# Near-Infrared Spectral Monitoring of Pluto's Ices: Spatial Distribution and Secular Evolution

W.M. Grundy[1,2], C.B. Olkin[2,3], L.A. Young[2,3],

M.W. Buie[2,3], and E.F. Young[2,3].

1. Lowell Observatory, 1400 W. Mars Hill Rd., Flagstaff AZ 86001.
2. Visiting or remote observer at the Infrared Telescope Facility, which is operated by the University of Hawaii under Cooperative Agreement no. NNX-08AE38A with the National Aeronautics and Space Administration, Science Mission Directorate, Planetary Astronomy Program.
3. Southwest Research Institute, 1050 Walnut St. #300, Boulder CO 80302.

## ABSTRACT

We report results from monitoring Pluto's 0.8 to 2.4 µm reflectance spectrum with IRTF/SpeX on 65 nights over the dozen years from 2001 to 2012. The spectra show vibrational absorption features of simple molecules $CH_4$, CO, and $N_2$ condensed as ices on Pluto's surface. These absorptions are modulated by the planet's 6.39 day rotation period, enabling us to constrain the longitudinal distributions of the three ices. Absorptions of CO and $N_2$ are concentrated on Pluto's anti-Charon hemisphere, unlike absorptions of less volatile $CH_4$ ice that are offset by roughly 90° from the longitude of maximum CO and $N_2$ absorption. In addition to the diurnal variations, the spectra show longer term trends. On decadal timescales, Pluto's stronger $CH_4$ absorption bands have been getting deeper, while the amplitude of their diurnal variation is diminishing, consistent with additional $CH_4$ absorption at high northern latitudes rotating into view as the sub-Earth latitude moves north (as defined by the system's angular momentum vector). Unlike the $CH_4$ absorptions, Pluto's CO and $N_2$ absorptions appear to be declining over time, suggesting more equatorial or southerly distributions of those species. Comparisons of geometrically-matched pairs of observations favor geometric explanations for the observed secular changes in CO and $N_2$ absorption, although seasonal volatile transport could be at least partly responsible. The case for a volatile transport contribution to the secular evolution looks strongest for $CH_4$ ice, despite it being the least volatile of the three ices.

# 1. Introduction

Reflectance spectroscopy has had a remarkable record of revealing the surface compositions of icy solar system bodies, ranging from satellites of the four giant planets to icy dwarf planets beyond Neptune's orbit (see reviews by Cruikshank et al. 1998a,b; Clark et al. 2013; and de Bergh et al. 2013; and references therein). In addition to nearly ubiquitous water ice, more exotic ices of methane, ethane, nitrogen, oxygen, hydrogen cyanide, ammonia, carbon monoxide, carbon dioxide, and sulfur dioxide have been identified on outer solar system bodies from their characteristic patterns of absorption bands imprinted on visible to infrared reflectance spectra. Several of these ices are highly volatile (e.g., Brown and Zeigler 1980; Fray and Schmitt 2009) and could be mobilized on seasonal timescales, even at the cryogenic temperatures prevalent in the outer solar system (Hansen and Paige 1996; Spencer et al. 1997; Trafton et al. 1998; Young 2012a). In addition to discovering the existence of such ices, spectroscopy can be a powerful tool for monitoring their evolution over time. This prospect is especially enticing for Pluto's surface, where volatile $N_2$, $CH_4$, and CO ices are all known to exist (Owen et al. 1993; Douté et al. 1999). Having passed through both perihelion and equinox within the past few decades, Pluto's surface and atmosphere are thought to be undergoing seasonal evolution at an especially rapid pace, at least compared with aphelion. Stellar occultations indicate that the atmospheric pressure has been increasing (Elliot et al. 2003; Sicardy et al. 2003; Young 2012b and references therein) and photometric lightcurves and Hubble Space Telescope imagery show that Pluto's surface albedo patterns are not static (Buratti et al. 2003; Schaefer et al. 2008; Buie et al. 2010a,b). NASA's New Horizons spacecraft will return a detailed, up-close look at the system in 2015 (e.g., Young et al. 2008). Understanding how that brief snapshot fits into the broader, seasonal context is crucial for getting the most scientific understanding out of the New Horizons flyby.

Various teams have been spectroscopically monitoring the Pluto system at near-infrared wavelengths to help constrain the spatial distribution and long term evolution of its surface ices (e.g., Rudy et al. 2003; Verbiscer et al. 2007; Merlin et al. 2010). From 1995 to 1998, we collected Pluto spectra on 83 nights at Lowell Observatory's 72" Perkins telescope (Grundy and Buie 2001). Over the past dozen years, our spectroscopic monitoring campaign has continued at NASA's InfraRed Telescope Facility (IRTF). This paper presents the ensemble of the IRTF data and discusses some patterns seen in the data. Companion papers will go into greater detail with radiative transfer models fitted to the Pluto spectra.

# 2. Data Acquisition and Reduction

Spectra were recorded during a series of observing runs from 2001 through 2012 at NASA's 3 m IRTF, located at an altitude of 4168 m on the summit of Mauna Kea. Most observing was done remotely by observers situated on the mainland as described by Bus et al. (2002), during scheduled telescope time allocations of typically two to four hours duration, although longer allocations were occasionally requested in order to obtain higher quality data. We used the short cross-dispersed mode of the SpeX spectrograph (Rayner et al. 1998, 2003). This mode divides the 0.8 to 2.4 µm wavelength range across five spectral orders, recorded on a $1024 \times 1024$ InSb detector array. Usable data were obtained on 65 nights, with circumstances described in Table 1. A few of the resulting spectra have been shown previously (Olkin et al. 2007; Protopapa et al. 2008; Tegler et al. 2012; plus several conference presentations), but this paper presents most of them for the first time, along with a full description of data acquisition and reduction procedures.

**Table 1**

Observational circumstances

| UT date of observation mean-time | Sky conditions[a] (on site or remote) | Slit width (″) | H band seeing (″) | Sub-Earth longitude[b] (°E) | Sub-Earth latitude[b] (°N) | Phase angle (°) | Pluto integration (min) |
|---|---|---|---|---|---|---|---|
| 2001/07/04 8ʰ.98 | Cirrus (HI) | 0.5 | 1.1 | 196.5 | 27.7 | 0.99 | 120 |
| 2001/07/07 9ʰ.20 | Cirrus (HI) | 0.5 | 1.0 | 26.9 | 27.6 | 1.06 | 147 |
| 2001/07/08 8ʰ.30 | Clear (HI) | 0.5 | 1.1 | 332.7 | 27.6 | 1.09 | 192 |
| 2002/06/24 6ʰ.54 | Mostly clear (HI) | 0.5 | 1.1 | 352.0 | 30.0 | 0.64 | 44 |
| 2002/06/25 6ʰ.30 | Cirrus (HI) | 0.5 | 0.6 | 296.2 | 29.9 | 0.66 | 46 |
| 2002/06/26 6ʰ.49 | Clear (HI) | 0.3 | 0.6 | 239.3 | 29.9 | 0.69 | 48 |
| 2002/07/15 7ʰ.24 | Clear (HI) | 0.3 | 0.6 | 246.8 | 29.5 | 1.19 | 60 |
| 2002/07/16 7ʰ.29 | Some clouds (HI) | 0.3 | 0.7 | 190.4 | 29.5 | 1.21 | 80 |
| 2002/07/17 6ʰ.94 | Partly cloudy (HI) | 0.3 | 0.8 | 134.8 | 29.4 | 1.23 | 96 |
| 2002/07/18 6ʰ.81 | Thin cirrus (HI) | 0.3 | 0.8 | 78.8 | 29.4 | 1.26 | 88 |
| 2002/07/19 7ʰ.06 | Thin cirrus (HI) | 0.3 | 0.7 | 21.8 | 29.4 | 1.28 | 78 |
| 2002/07/20 6ʰ.80 | Thin cirrus (HI) | 0.3 | 0.8 | 326.1 | 29.4 | 1.30 | 92 |
| 2002/07/21 6ʰ.81 | Thin cirrus (HI) | 0.3 | 0.6 | 269.7 | 29.4 | 1.33 | 96 |
| 2002/07/22 6ʰ.72 | Clear (HI) | 0.3 | 0.7 | 213.6 | 29.4 | 1.35 | 104 |
| 2003/08/09 6ʰ.82 | Clear (r) | 0.5 | 0.6 | 225.0 | 31.1 | 1.63 | 96 |
| 2005/04/15 13ʰ.15 | Thin cirrus (r) | 0.3 | 0.5 | 103.0 | 37.1 | 1.59 | 176 |
| 2005/05/09 13ʰ.06 | Mostly clear (r) | 0.3 | 0.7 | 190.7 | 36.8 | 1.10 | 188 |
| 2005/05/23 11ʰ.27 | Heavy cirrus (r) | 0.3 | 0.7 | 125.9 | 36.5 | 0.73 | 184 |
| 2005/05/24 12ʰ.03 | Clear (r) | 0.3 | 0.7 | 67.8 | 36.5 | 0.70 | 268 |
| 2006/07/19 8ʰ.32 | Clear (r) | 0.3 | 0.8 | 106.7 | 37.0 | 1.01 | 148 |
| 2006/07/24 7ʰ.83 | Heavy cirrus (r) | 0.3 | 1.5 | 186.1 | 37.0 | 1.14 | 100 |
| 2006/07/25 7ʰ.73 | Clear (r) | 0.3 | 1.0 | 130.0 | 36.9 | 1.16 | 160 |
| 2006/07/26 7ʰ.71 | Clear (r) | 0.3 | 0.9 | 73.7 | 36.9 | 1.19 | 168 |
| 2006/08/08 7ʰ.13 | Thin cirrus (r) | 0.3 | 0.5 | 62.4 | 36.7 | 1.47 | 126 |
| 2006/08/25 6ʰ.80 | Clear (r) | 0.3 | 0.7 | 185.1 | 36.5 | 1.72 | 136 |
| 2007/06/23 10ʰ.78 | Clear (r) | 0.3 | 0.6 | 72.0 | 39.5 | 0.26 | 32 |
| 2007/06/25 10ʰ.66 | Clear (r) | 0.3 | 0.5 | 319.5 | 39.4 | 0.30 | 36 |
| 2007/08/06 7ʰ.27 | Clear (r) | 0.5 | 0.7 | 120.7 | 38.5 | 1.36 | 104 |
| 2007/08/07 7ʰ.50 | Clear (r) | 0.5 | 0.6 | 63.8 | 38.5 | 1.39 | 80 |
| 2007/08/08 7ʰ.06 | Clear (r) | 0.5 | 0.6 | 8.5 | 38.5 | 1.41 | 108 |
| 2007/08/09 6ʰ.90 | Clear (r) | 0.5 | 0.6 | 312.5 | 38.5 | 1.42 | 112 |
| 2007/08/10 7ʰ.07 | A little cirrus (r) | 0.5 | 0.6 | 255.7 | 38.5 | 1.44 | 116 |
| 2007/08/11 7ʰ.65 | Clear (r) | 0.5 | 0.7 | 198.0 | 38.5 | 1.46 | 124 |
| 2008/06/16 10$^h$.99 | A little cirrus (r) | 0.5 | 0.5 | 355.4 | 41.4 | 0.25 | 52 |
| 2008/06/17 10$^h$.94 | Clear (r) | 0.5 | 0.5 | 299.1 | 41.4 | 0.23 | 64 |
| 2008/06/18 11$^h$.10 | Cirrus (r) | 0.5 | 0.6 | 242.4 | 41.3 | 0.22 | 34 |
| 2008/06/19 10$^h$.63 | Cirrus (r) | 0.5 | 0.5 | 187.2 | 41.3 | 0.21 | 84 |
| 2008/06/20 10$^h$.67 | Clear (r) | 0.5 | 0.6 | 130.7 | 41.3 | 0.21 | 60 |
| 2008/06/21 10$^h$.68 | Clear (r) | 0.5 | 0.5 | 74.3 | 41.3 | 0.21 | 60 |
| 2009/06/18 12$^h$.52 | Clear (r) | 0.5 | 0.8 | 184.8 | 43.0 | 0.24 | 64 |
| 2009/06/19 12$^h$.14 | Clear, humid (r) | 0.5 | 1.1 | 129.3 | 43.0 | 0.22 | 80 |
| 2009/06/20 12$^h$.84 | Clear, humid (r) | 0.5 | 0.9 | 71.3 | 43.0 | 0.20 | 84 |
| 2009/06/21 12$^h$.40 | Clear (r) | 0.5 | 1.1 | 16.0 | 43.0 | 0.19 | 58 |
| 2009/06/22 12$^h$.62 | Clear (r) | 0.5 | 0.7 | 319.1 | 42.9 | 0.19 | 88 |
| 2010/05/08 14$^h$.11 | Clear (r) | 0.5 | 0.6 | 277.1 | 45.5 | 1.34 | 80 |
| 2010/05/28 14$^h$.01 | Some clouds (r) | 0.5 | 1.1 | 230.2 | 45.1 | 0.86 | 64 |
| 2010/06/24 10$^h$.52 | Clear (r) | 0.5 | 0.7 | 157.0 | 44.5 | 0.17 | 92 |
| 2010/06/25 10$^h$.25 | Clear (r) | 0.5 | 0.5 | 101.3 | 44.5 | 0.16 | 124 |
| 2010/06/30 10$^h$.22 | Fog (r) | 0.5 | 0.7 | 179.6 | 44.4 | 0.22 | 96 |
| 2010/07/01 9$^h$.99 | Clear (r) | 0.5 | 0.7 | 123.8 | 44.4 | 0.24 | 112 |
| 2011/05/22 14$^h$.77 | Clear (r) | 0.5 | 0.7 | 152.1 | 46.8 | 1.07 | 52 |
| 2011/06/01 13$^h$.96 | Clear (r) | 0.3 | 0.8 | 310.5 | 46.6 | 0.81 | 72 |
| 2011/06/02 13$^h$.89 | Trace of clouds (r) | 0.3 | 0.7 | 254.3 | 46.6 | 0.78 | 76 |
| 2011/06/06 14$^h$.43 | Clear (r) | 0.3 | 0.6 | 27.6 | 46.5 | 0.67 | 64 |
| 2011/06/07 13$^h$.91 | Clear (r) | 0.3 | 0.6 | 332.5 | 46.5 | 0.64 | 40 |
| 2011/06/11 13$^h$.82 | Clear (r) | 0.3 | 0.5 | 107.3 | 46.4 | 0.53 | 52 |
| 2011/06/12 11$^h$.90 | Partly cloudy (r) | 0.3 | 0.4 | 55.5 | 46.4 | 0.50 | 64 |
| 2011/07/15 9$^h$.02 | Clear (r) | 0.3 | 0.9 | 2.8 | 45.6 | 0.54 | 64 |
| 2012/05/12 13$^h$.51 | Clear (r) | 0.3 | 0.5 | 247.7 | 48.4 | 1.32 | 104 |
| 2012/05/13 13$^h$.83 | Clear (r) | 0.3 | 0.6 | 190.6 | 48.4 | 1.30 | 120 |
| 2012/05/14 13$^h$.92 | Trace of cirrus (r) | 0.3 | 0.7 | 134.0 | 48.4 | 1.28 | 92 |
| 2012/05/15 14$^h$.04 | Clear (r) | 0.3 | 0.5 | 77.3 | 48.4 | 1.25 | 120 |
| 2012/05/16 14$^h$.03 | Clear (r) | 0.3 | 0.6 | 21.0 | 48.4 | 1.23 | 96 |
| 2012/05/17 13$^h$.97 | Clear (r) | 0.3 | 0.6 | 324.8 | 48.4 | 1.21 | 104 |
| 2012/09/09 6$^h$.19 | Clear (r) | 0.3 | 0.7 | 342.9 | 46.3 | 1.68 | 40 |

Table notes:

[a.] On-site observing runs are indicated by “(HI)” and remote observing runs are indicated by “(r)”.

[b.] Throughout this paper, geometry on Pluto is defined as in previous publications of this author, using a right-hand-rule coordinate system in which North is the direction of Pluto's spin vector and East is the direction of sunrise. Zero longitude is defined by the sub-Charon point. Longitudes and latitudes tabulated here were computed by assuming Pluto's spin state is tidally locked to the Buie et al. (2012) orbit of Charon.

On each night, sets of Pluto observations lasting from 30 minutes up to an hour were sandwiched between sets of nearby solar analog star observations. Each set of observations, star or Pluto, consisted of at least 3 spectra taken in an "A" beam position and at least 3 more taken in a separate "B" beam position, interleaved according to an ABBAAB... pattern. Star integration times were set to a minimum of 30 seconds to ensure that the spatial profile in each star integration was averaged over numerous telescope guiding updates, in order to be as similar as possible to the longer 2 minute Pluto integrations, while still leaving the majority of available time for Pluto integrations. In cases where the star was too bright to integrate for 30 seconds without saturating, multiple, consecutive, shorter integrations were combined through co-addition to synthesize 30 second integrations. A and B spectral images were subtracted pairwise to remove most telluric sky emission features. Spectral extraction of the A and B components from each such subtracted pair was accomplished using the Horne (1986) optimal extraction algorithm. Details of our implementation have been described previously and, for the sake of brevity, interested readers are referred to Grundy et al. (2010) and references therein. However, it is worth mentioning here that the spatial profiles used to extract Pluto spectra were constructed from the bracketing star observations, rather than from the Pluto frames themselves.

Pluto's motion along its heliocentric orbit required a new nearby solar analog star to be selected every few years. Stars used for this purpose (and the years they were used) were SAO 160066 (2001), HD 153631 (2003), HD 160788 (2005-2009), and HD 170379 (2010-2012). Spectra of these stars were compared with additional observations of established solar analogs 16 Cyg B, BS 5968, BS 6060, SA 112-1333, Hyades 64, and SA 105-56. The nearby solar analogs were found to be spectrally consistent with these better known analogs, with the exception of HD 170379, which was found to have an effective temperature of ~6500 K, somewhat hotter than the Sun. To account for the temperature difference, a correction based on the ratio of two Planck functions was applied to observations using that star. Pluto spectra were divided by spectra of bracketing nearby solar analogs to remove most instrumental and telluric effects. During 2002 we tried a different procedure of observing an ensemble of well known solar analogs over a broad range of airmasses to construct an extinction model for the night. After correcting all Pluto and star spectra to a common airmass, the Pluto average was divided by the star average. This procedure was hampered by wavelength shifts from instrument flexure, leading us to rely on a nearby solar analog in all subsequent years. Because the observed Pluto flux is entirely due to reflected sunlight over this wavelength range, the Pluto/solar analog ratio produced spectra proportional to the reflectance of Pluto. It is not possible to quote absolute albedos from our narrow-slit spectra, since variable slit losses (e.g., due to changes in guiding, seeing, or focus) undermine the photometric fidelity of comparisons between targets and standards. Residual telluric features remain near 1.4 and 1.9 µm, where strong and narrow $H_2O$ vapor absorptions make sky transparency especially variable in time. Example spectra from some representative nights are plotted in Fig. 1, showing how data quality varies from relatively short total integration times, such as the 2007 June 23 spectrum, to much longer total integrations with multiple interleaved Pluto and star observations as on 2005 May 24.

We selected one of two different slit widths for use during a night's observations. An 0.5″ slit provided a reasonable match to near-infrared seeing at Mauna Kea, resulting in a typical spectral resolving power of $R$ = 1200 (wavelength divided by measured full width at half maximum for an unresolved line). A narrower 0.3″ slit boosted the resolution to around $R$ = 1900, at a cost of increased slit losses. SpeX's internal integrating sphere, illuminated by a low-pressure argon arc lamp, was observed at similar sky positions as Pluto so as to duplicate instrument flexure conditions experienced during the Pluto observations. A quadratic wavelength calibration curve was derived for each spectral order from the ensemble of arc lamp spectra obtained over the course of the Pluto observations. For additional protection against flexure effects, the derived wavelengths of telluric sky emission lines extracted from the Pluto frames themselves were checked against their published wavelengths (Rousselot et al. 2000; Hanuschik 2003). Occasionally, this check called for a shift in the constant term of the wavelength calibration relative to what was measured for the argon spectra. Spectral profiles of both arc lamp and sky emission lines were well approximated by Gaussians having full width at half maximum (FWHM) of ~2 pixels for the 0.3″ slit and ~3 pixels for the 0.5″ slit. Wavelength uncertainty, primarily due to flexure and to a few sparse regions in the argon spectrum, is generally less than a pixel. Flexural shifts over the course of a night's observations degraded the spectral resolution of final nightly average spectra, especially for nights with longer total integration times. On such nights, measured resolutions of $R \approx 1600$ and 1100 were more typical for the 0.3″ and 0.5″ slits, respectively.

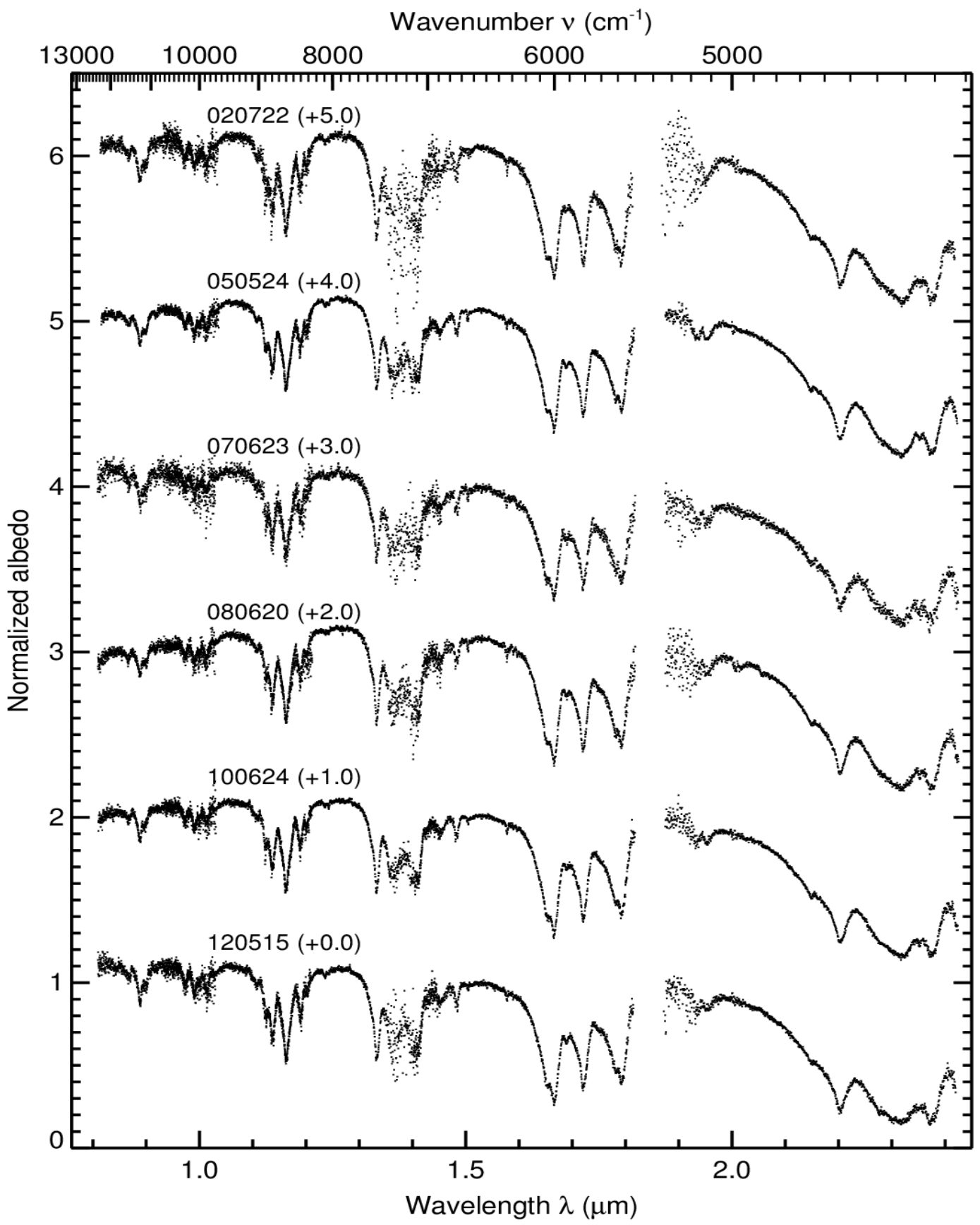

**Figure 1.** Example spectra obtained on UT dates indicated in YYMMDD format with vertical offsets indicated in parentheses. Noisy areas or gaps around 1.0, 1.4, and 1.9 µm coincide with higher opacity in the terrestrial atmosphere.

As seen from Earth, Pluto and its largest satellite Charon are never separated by more than about 1″ on the sky, so light from both objects can enter the spectrograph slit and be blended in our spectra. Unfortunately, the amount of light received from Charon is highly variable, depending on the separation between the two objects, seeing conditions, slit width, and the orientation of the slit relative to the Pluto-Charon position angle. At higher airmasses we would try to orient the 15″ long slit near the parallactic angle to minimize effects of differential refraction by the Earth's atmosphere, while at lower airmasses we would try to place it along the Pluto-Charon line. However, over the past decade Pluto has been passing through a highly crowded region near the galactic center, so the need to keep background sources out of the slit often left us with little choice of slit orientation. Our use of point-source spatial profiles based on the star observations helped reduce the Charon contribution on nights of better seeing, but on poor seeing nights, signal from the two bodies was highly blended. Charon accounts for about a fifth of the total surface area in the Pluto system so it could contribute up to a fifth of the observed signal, if its albedo was the same as Pluto's. Charon's albedo is actually lower than Pluto's at most wavelengths between 0.8 and 2.4 µm (Fink and DiSanti 1988; Douté et al. 1999; Buie & Grundy 2000; Grundy and Buie 2002), diminishing its potential contribution to our spectra. But in narrow spectral regions centered about 1.72 and 1.79 µm, and also in the range from around 2.18 to 2.4 µm, strong $CH_4$ ice absorptions reduce Pluto's albedo to below that of Charon, so these spec-tral regions are likeliest to show effects of variable Charon contamination. Spectral contributions from Nix, Hydra, and the other small satellites are negligible, both because they are further away from Pluto and because they are much fainter (e.g., Weaver et al. 2006; Showalter et al. 2011, 2012).

Pluto's disk as seen from Earth is an unresolved 0.1″ across, so each of our observations represents the reflectance integrated over whichever hemisphere of Pluto is oriented toward Earth at the time of the observation. The region near the center of the disk contributes most strongly to the observed signal, with foreshortened regions approaching the limb contributing less. The sub-Earth longitudes and latitudes of our 65 observations are shown in Fig. 2. The full data set provides relatively dense sampling as a function of sub-Earth longitude, making it particularly well-suited to investigating the longitudinal distributions of Pluto's volatile CO, $CH_4$, and $N_2$ ices, as discussed in the next section.

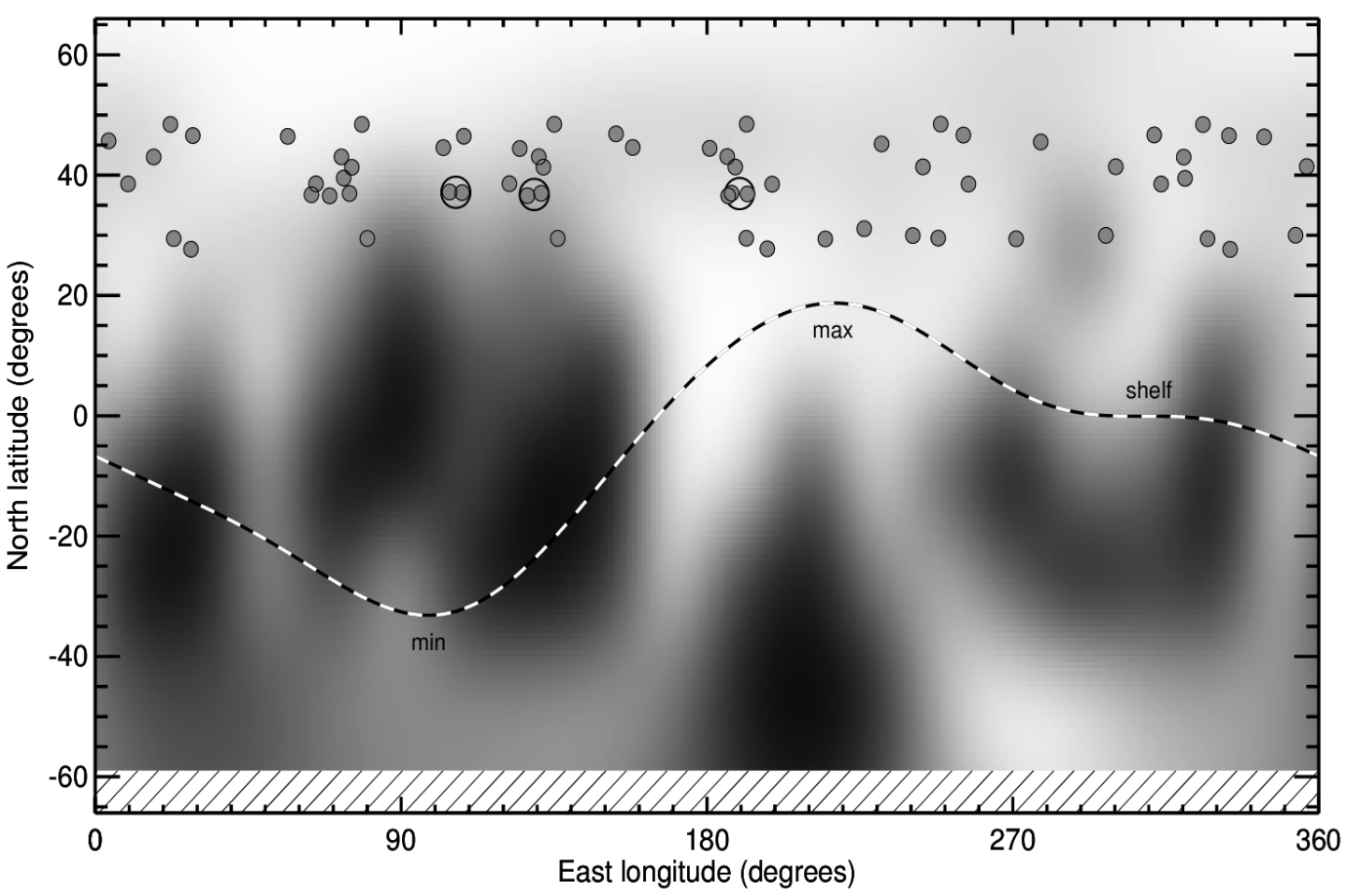

**Figure 2.** Distribution of sub-Earth longitudes and latitudes on Pluto at the times of our IRTF observations, superimposed on a map of single scattering albedos at visible wavelengths from Buie et al. (2010b), based on Hubble Space Telescope observations obtained between 2002 June and 2003 June. The sub-Charon point is at 0° E, 0° N and the anti-Charon point is at 180° E, 0° N. The hatched region was not illuminated by the Sun at the time of the Buie et al. observations and thus was not mapped. Pluto's visible lightcurve from the same HST observations (Buie et al. 2010a) is shown as a dashed curve, arbitrarily scaled, with the main features labeled. Circled pairs of observations are discussed in Section 4.

# 3. Longitudinal Distributions of Ices

## *3.1 Carbon Monoxide Ice*

Two narrow CO ice absorption bands are visible in our spectra. Both are overtones of a fundamental vibrational transition at 4.67 µm (Quirico and Schmitt 1997a). Despite its being the stronger of the two, the 2-0 overtone at 2.35 µm is difficult to quantify in our spectra because it falls between two very strong and overlapping $CH_4$ ice absorptions. The 3-0 overtone at 1.58 µm occupies a much less cluttered spectral region. A simple measure of an absorption band's strength is its equivalent width. This was computed for the 3-0 CO band in each of our spectra by integrating one minus the ratio of spectrum/continuum over wavelengths between 1.575 and 1.581 µm, with wavelengths from 1.571 to 1.575 µm and from 1.581 to 1.585 µm being used to constrain a linear continuum model. Based on near-infrared spectral observations of this same band obtained during 1995 to 1998, Grundy and Buie (2001) reported that CO ice absorption on Pluto's anti-Charon hemisphere was much stronger than on the sub-Charon face. The new data independently confirm that result, as shown in Fig. 3. Comparing with the map in Fig. 2, a high-albedo region centered around 180° E, 10° N could plausibly be the site of the excess CO ice absorption. CO ice absorption appears greatly reduced on Pluto's Charon-facing hemisphere, but it is definitely not zero there. Averaging the thirteen observations obtained within 40° of zero lon-gitude gives a securely non-zero equivalent width of 0.076 ± 0.006 nm.

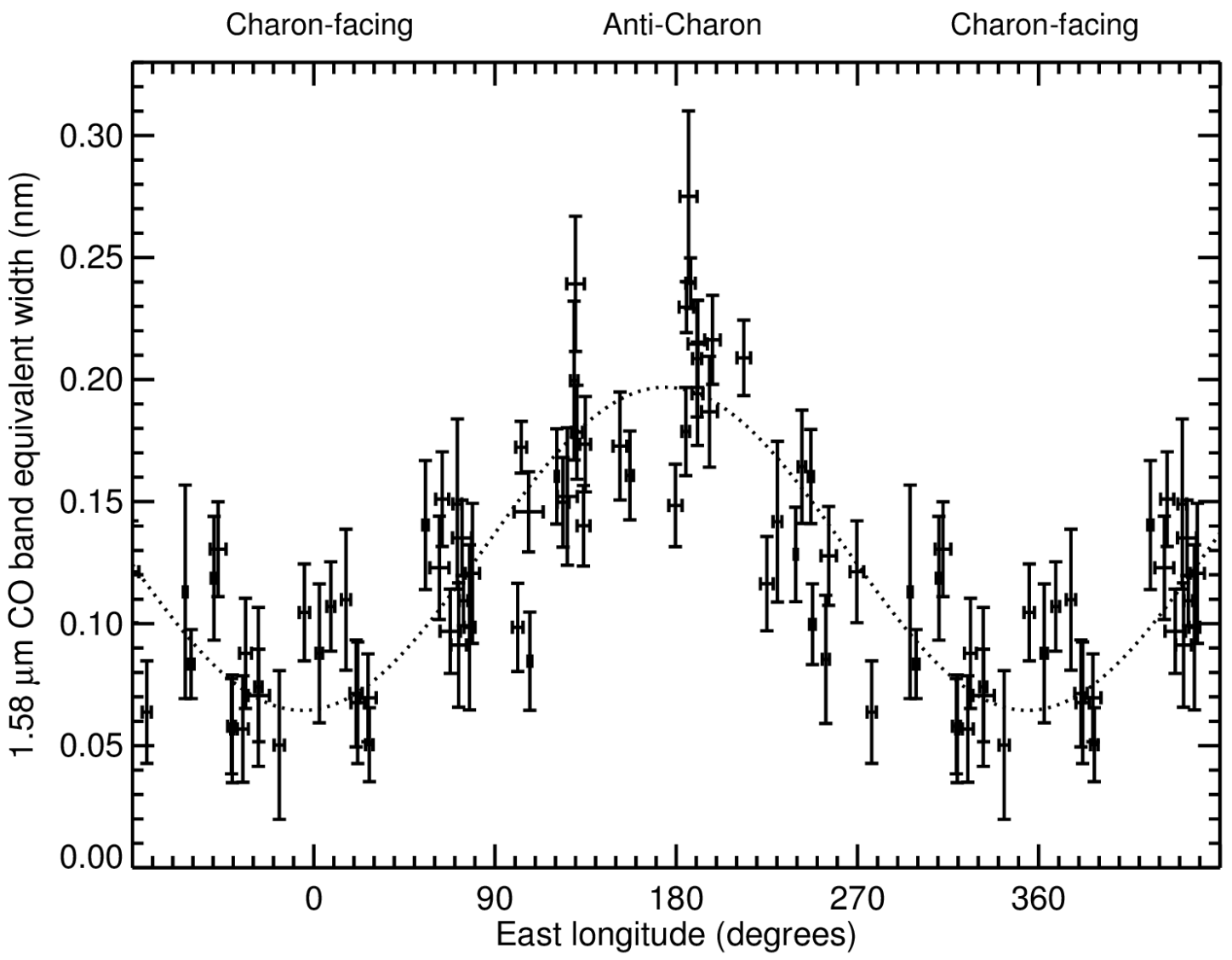

**Figure 3.** Equivalent width of the 1.58 µm 3-0 CO ice absorption band, showing a roughly three-fold variation as Pluto rotates, with maximum absorption being centered on the anti-Charon hemisphere. Each point represents data from one night's observations, with the horizontal error bar indicating the duration of the observations and the vertical error bar indicating the 1-σ uncertainty in the equivalent width of the 1.58 µm CO ice band. Points are duplicated outside the 0° to 360° longitude range to better show the cyclical pattern. The dotted curve is a sinusoid fitted to the data, with a constant offset, amplitude, and phase of maximum absorption as free parameters. The fitted values are 0.13 nm, 0.066 nm, and 174°, respectively.

## 3.2 Methane Ice

The four fundamental vibrational modes of the $CH_4$ molecule produce a complex array of overtone and combination absorption bands distributed across Pluto's near-infrared spectrum. Indeed, Pluto has so many overlapping $CH_4$ absorption bands that computing equivalent widths for individual bands is not feasible. So instead we computed fractional depths by subtracting from unity the ratio of the spectral signal at the center of a $CH_4$ band to the associated continuum signal. Where continuum wavelengths were available on both sides of a band, a linear continuum model was computed from the two, and evaluated at the wavelength of the band canter. Otherwise a simple ratio was computed. The spectral signals in band centers were obtained by averaging over a small wavelength interval to overcome noise and effects of variable spectral resolution. Examples of this fractional band depth are shown for four bands in Fig. 4, and wavelengths used in computing them (for these and eight other $CH_4$ bands) are listed in Table 2. Sinusoids fitted to the fractional band depths show trends that are generally similar from one band to the next, with maximum absorption in the 260° to 320° E longitudes range, and minimum absorption in the 80° to 140° E range.

**Table 2**

$CH_4$ ice absorption bands, wavelength windows, and sinusoidal fits to fractional band depths

| Absorption band | Wavelength windows (μm) | | | Sinusoidal fit | | |
|---|---|---|---|---|---|---|
| | 1st continuum | Band | 2nd continuum | Const. | Ampl. | Lon. (°E) |
| 0.89 μm ($2\nu_1 + \nu_3 + 2\nu_4$) | 0.845 - 0.857 | 0.8878 - 0.8902 | 0.920 - 0.932 | 0.20 | 0.039 | 315 |
| 0.97 μm ($2\nu_3 + \nu_4$) | 0.930 - 0.950 | 0.9710 - 0.9750 | 1.050 - 1.070 | 0.11 | 0.029 | 319 |
| 0.99 μm ($\nu_1 + 2\nu_3 + \nu_4$) | 0.930 - 0.950 | 0.9895 - 0.9925 | 1.050 - 1.070 | 0.16 | 0.035 | 322 |
| 1.14 μm ($\nu_2 + 2\nu_3 + \nu_4$) | 1.060 - 1.080 | 1.1350 - 1.1390 | 1.245 - 1.265 | 0.39 | 0.032 | 292 |
| 1.16 μm ($2\nu_1 + \nu_2 + \nu_4$) | 1.060 - 1.080 | 1.1605 - 1.1645 | 1.245 - 1.265 | 0.49 | 0.032 | 279 |
| 1.19 μm ($\nu_1 + \nu_3 + 2\nu_4$) | 1.060 - 1.080 | 1.1880 - 1.1910 | 1.245 - 1.265 | 0.28 | 0.028 | 301 |
| 1.33 μm ($\nu_2 + 2\nu_3$) | 1.250 - 1.280 | 1.3305 - 1.3345 | 1.510 - 1.540 | 0.48 | 0.032 | 262 |
| 1.66 μm ($2\nu_2 + \nu_3$) | 1.510 - 1.540 | 1.6635 - 1.6675 | 1.980 - 2.010 | 0.69 | 0.036 | 257 |
| 1.72 μm ($\nu_2 + \nu_3 + \nu_4$) | 1.510 - 1.540 | 1.7185 - 1.7225 | 1.980 - 2.010 | 0.61 | 0.035 | 262 |
| 1.79 μm ($\nu_3 + 2\nu_4$) | 1.510 - 1.540 | 1.7910 – 1.7940 | 1.980 - 2.010 | 0.58 | 0.043 | 257 |
| 2.20 μm ($\nu_2 + \nu_3$) | 1.980 – 2.010 | 2.2005 - 2.2055 | n/a | 0.74 | 0.038 | 256 |
| 2.32 μm ($\nu_3 + \nu_4$) | 1.980 – 2.010 | 2.3130 - 2.3230 | n/a | 0.84 | 0.033 | 260 |

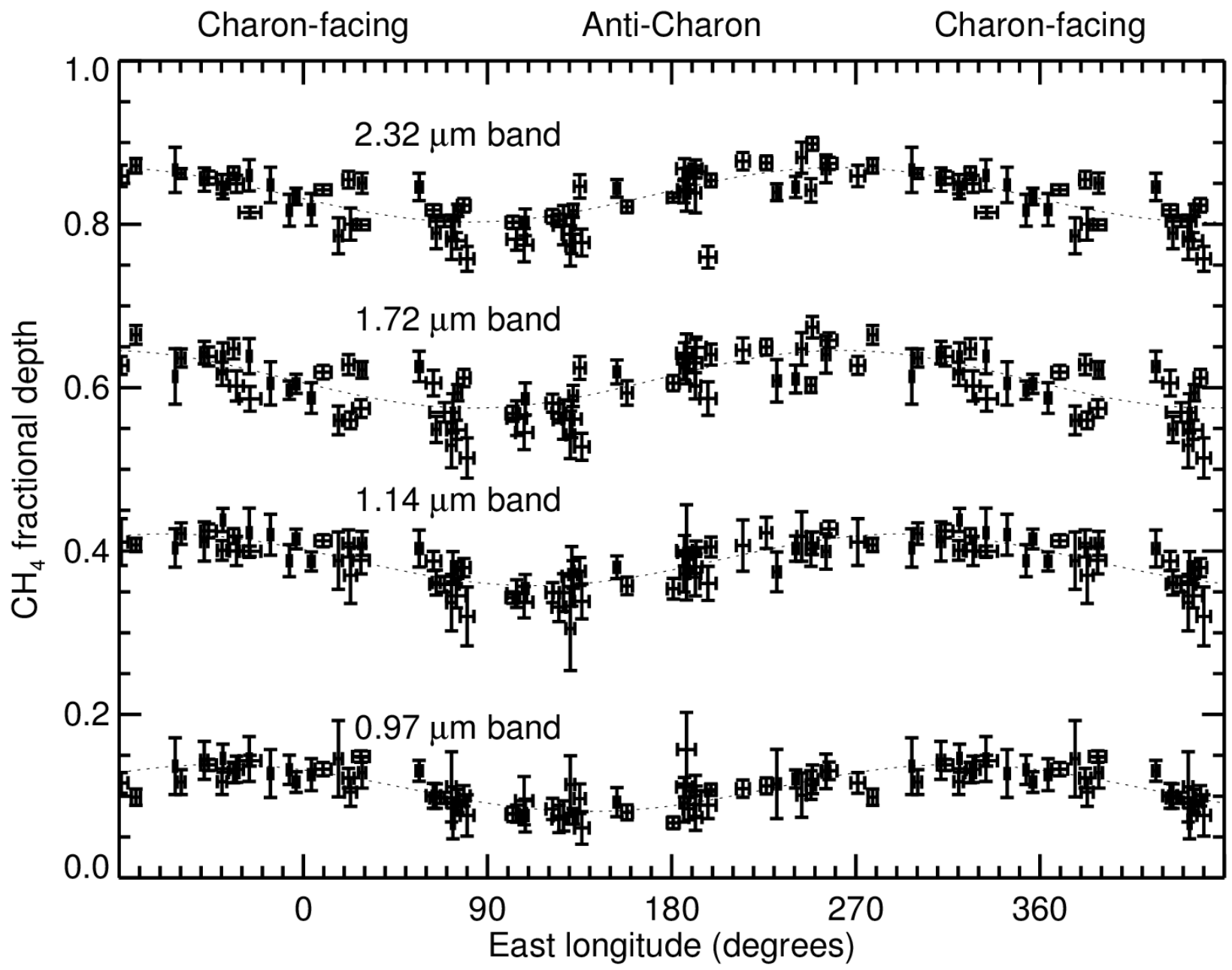

**Figure 4.** Fractional depths of four $CH_4$ ice absorption bands as a function of sub-Earth longitude on Pluto. As in Fig. 3, each point represents data from one night's observations, with the horizontal error bar indicating the duration of the observations and the vertical error bar indicating the 1-σ uncertainty in the measured depth of the indicated band. Sinusoidal fits are shown with dotted lines, with parameter values listed in Table 2. The larger scatter in fractional depths of the deeper bands is mostly due to secular evolution, as discussed in Section 4, but the stronger bands are also more vulnerable to variable contamination by Charon light.

Correlation between Pluto's near-infrared $CH_4$ band depths and visible lightcurve has been noted long ago (e.g., Buie and Fink 1987; Marcialis and Lebofsky 1991), and is also evident in our data. This longitudinal pattern is strikingly different from the Charon-centered geometry of the CO ice absorption. Correlation between Pluto's $CH_4$ band depths and the visual lightcurve supports a correlation between $CH_4$ absorption and albedo that has been suggested previously (e.g., Grundy and Buie 2001). But the IRTF data show subtle differences from one $CH_4$ band to the next. Plotting the longitude where each $CH_4$ band shows its maximum absorption against the mean fractional depth of each band (Fig. 5) reveals a strong correlation, with deeper bands showing maximum absorption near 260° E and shallower bands showing their maximum absorption further east, toward the Charon-facing hemisphere. This contrast between Pluto's weaker and stronger $CH_4$ bands has also been noted previously (e.g., Grundy and Fink 1996; Grundy and Buie 2001) and interpreted as indicating that the $CH_4$ responsible for producing the deeper $CH_4$ bands is more prevalent at longitudes corresponding to Pluto's lightcurve maximum, while terrains with the larger particle sizes or greater $CH_4$ enrichment needed to boost the depths of weaker $CH_4$ absorptions occur more on the Charon-facing hemisphere.

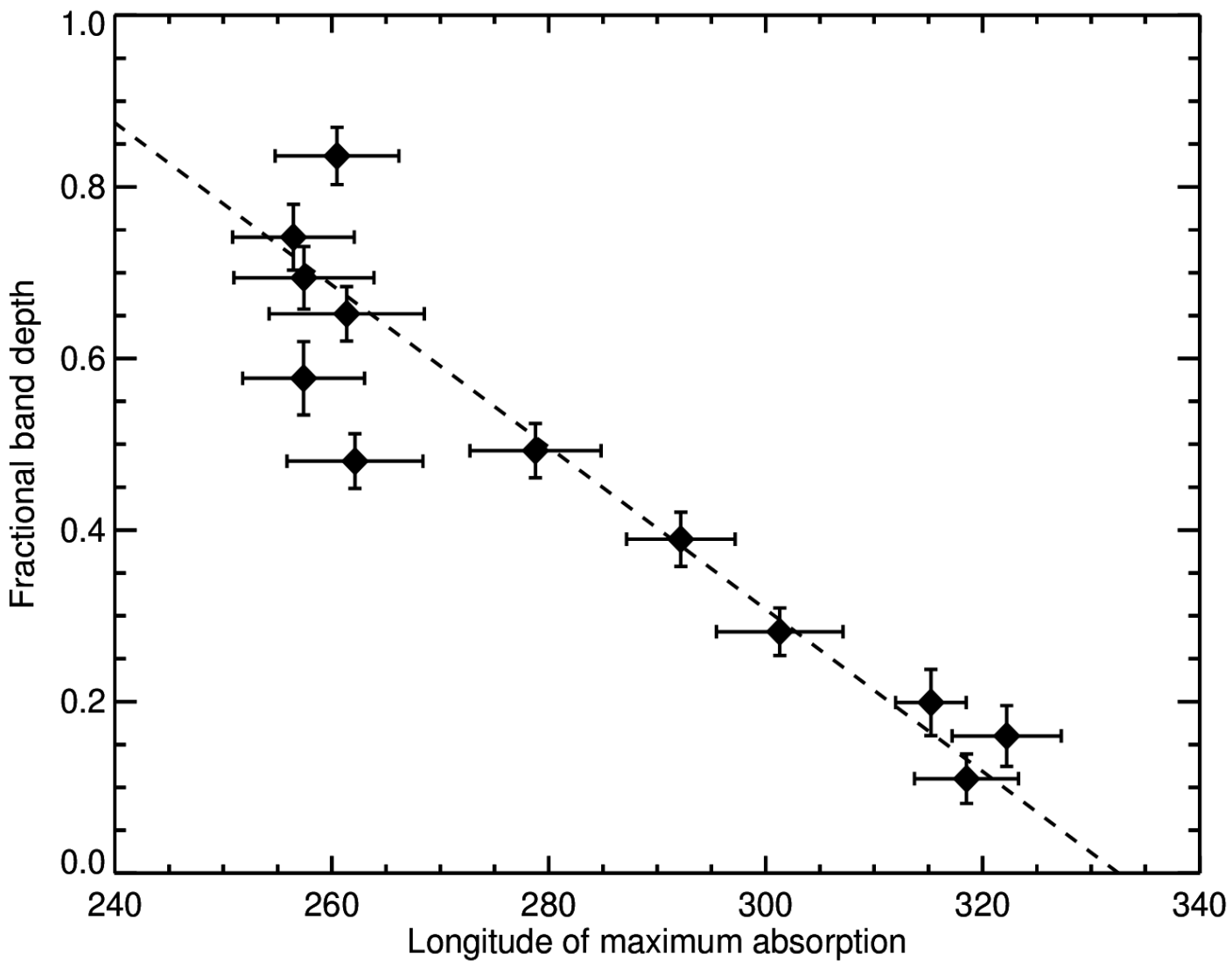

**Figure 5.** Comparison of the longitude of maximum absorption (determined from sinusoidal fits as shown in Fig. 4) and average fractional band depth, for the dozen methane bands listed in Table 2. Horizontal bars represent the 1-σ uncertainty in the maximum absorption longitude and vertical bars indicate the peak-to-peak amplitude of the sinusoidal fit. A linear regression is shown with a dashed line.

### *3.3 Nitrogen Ice*

$N_2$ is a non-polar molecule, so vibrational absorptions are not easily excited by incident photons. However, when the molecule's electron distribution is temporarily distorted by interaction with a neighboring molecule, a weak "collisionally induced" fundamental absorption appears at 4.25 µm with an even weaker overtone in our wavelength range at 2.15 µm (e.g., Shapiro and Gush 1966; Owen et al. 1993; Clark et al. 2013). This band appears as a subtle feature just to the left of a much deeper $CH_4$ ice band at 2.2 µm (see Fig. 1). The $N_2$ absorption is difficult to disentangle from the wings of the $CH_4$ band, requiring detailed modeling for an accurate estimation of the $N_2$ abundance (to be presented in a separate paper). But we can get a crude idea of the longitudinal distribution of $N_2$ absorption by computing an equivalent width, as we did for CO ice. Here we used wavelengths from 2.115 to 2.130 µm and from 2.165 to 2.180 µm to compute a linear continuum model and integrated one minus spectrum/continuum from 2.130 to 2.165 µm. The result, plotted in Fig. 6, has considerable scatter, but shows greater $N_2$ absorption on the anti-Charon hemisphere, similar to the pattern seen for CO ice, so we can tentatively associate both ices with the bright anti-Charon spot in the Buie et al. (2010b) albedo map. That $N_2$ and CO ices would exhibit similar spatial distributions is not surprising, since the two ices have similar volatilities (Brown and Ziegler 1980; Fray and Schmitt 2009) and are completely miscible in one another (Vetter et al. 2007). The absorptions of these two ices have also been observed to have very similar longitudinal distributions to one another on Triton, Neptune's large, captured satellite (Grundy et al. 2010).

Another way to investigate Pluto's $N_2$ ice is via the much more numerous $CH_4$ ice bands. This is possible because dilution of $CH_4$ in $N_2$ ice has the effect of blue-shifting the $CH_4$ bands relative to their positions in pure $CH_4$ ice (Quirico and Schmitt 1997b). Larger blue shifts correspond to greater dilution factors (Brunetto et al. 2008). An easy way to quantify the shift is by cross-correlation relative to a model for pure $CH_4$ ice, as described by Tegler et al (2008). The results are shown for four methane bands in Fig. 7. These (and other) $CH_4$ bands all show maximum blue shifts when the anti-Charon hemisphere is oriented toward Earth and minimum blue shifts when the sub-Charon hemisphere is visible, consistent with the $N_2$ distribution implied by the equivalent width of the $N_2$ band.

As Fig. 7 shows, different $CH_4$ absorption bands have different average blue shifts in wavelength units. These different bands also probe different average depths into the surface of Pluto, with more strongly absorbing bands sampling, on average, less deep within a particulate surface than weaker bands do. This fact has been used to search for depth-dependent $CH_4/N_2$ ratios by comparing shifts versus band depth (e.g., Licandro et al. 2006a,b; Tegler et al. 2008; Abernathy et al. 2009; Merlin et al. 2009, 2010). The not-always consistent results of these efforts point to an additional com-plication: $CH_4$ is not soluble in $N_2$ at all concentrations. When the solubility limit is exceeded, two separate phases co-exist, one $CH_4$-rich, and the other $N_2$-rich, as described by the binary phase diagram (Prokhvatilov and Yantsevich 1983; Lunine and Stephenson 1985). $CH_4$ bands in the $N_2$-rich phase exhibit blue shifts like those reported by Quirico and Schmitt (1997b) while $CH_4$ bands in the $CH_4$-rich phase exhibit much smaller blue shifts that have yet to be measured in the laboratory in a systematic way, although that work is now in progress. What the cross-correlation technique treats as a single shifted band should actually be modeled as a superposition of two distinct absorption bands. Approximating the $CH_4$-rich phase with pure $CH_4$ optical constants from Grundy et al. (2002) and the $N_2$-rich phase with a shifted version of the same, Tegler et al. (2010, 2012) fitted a two-phase model to a series of both weak and strong $CH_4$ bands, finding no convincing evidence for compositional trends with band depth. Applying a model like that to this new data set will enable much more quantitative compositional conclusions as a function of longitude and is the subject of a separate paper.

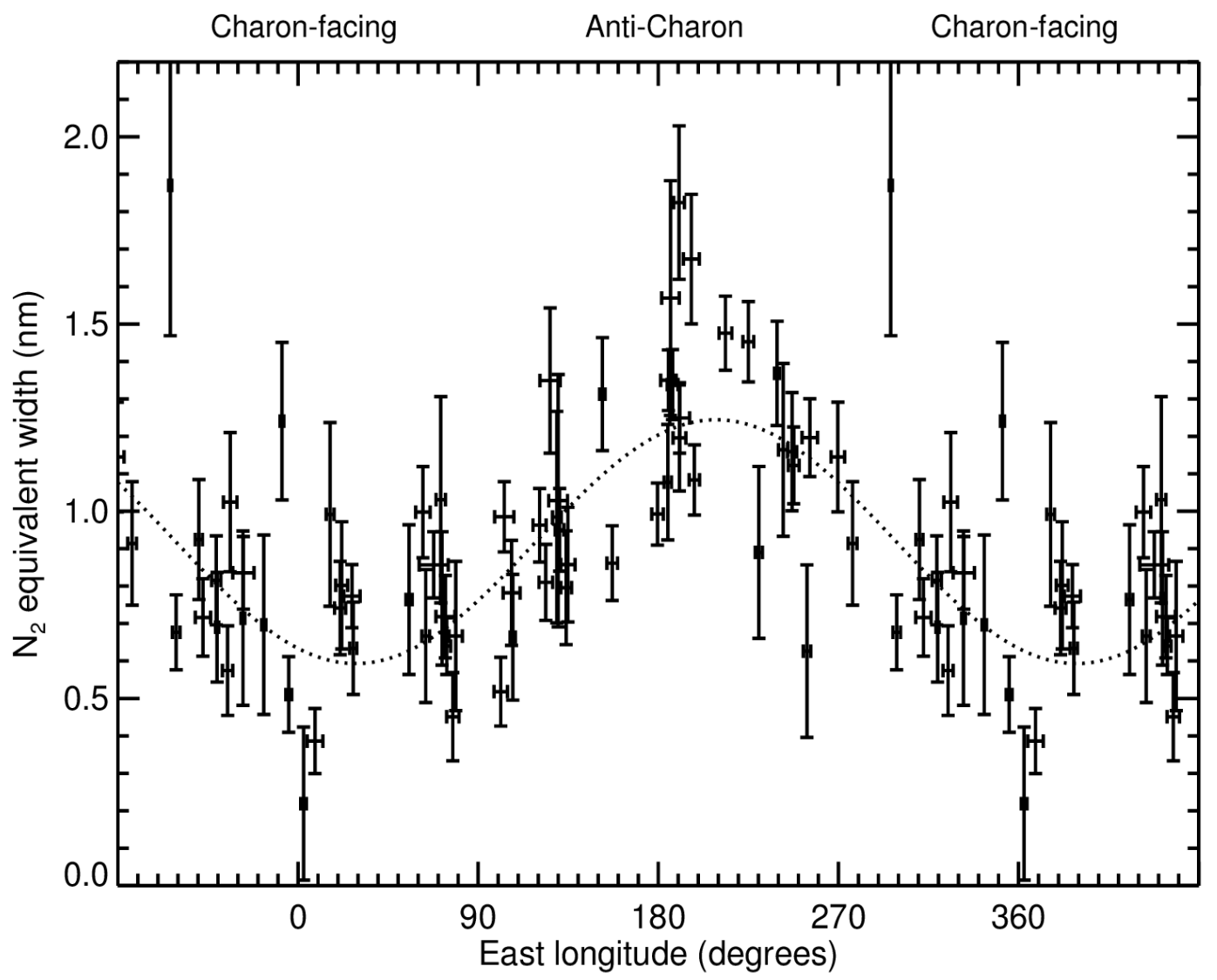

**Figure 6.** Approximate equivalent width of the 2.15 µm $N_2$ ice band as a function of sub-Earth longitude. Our use of a linear continuum approximation over wavelengths where there is clearly some curvature due to the wings of the strong 2.2 µm $CH_4$ band can be expected to cause systematic underestimation of the equivalent width, but at least it gives some idea of the longitude dependence. The dotted curve is a sinusoid fitted to the data, with a constant offset, amplitude, and phase as free parameters. The fitted values are 0.92 nm, 0.33 nm, and 209°, respectively.

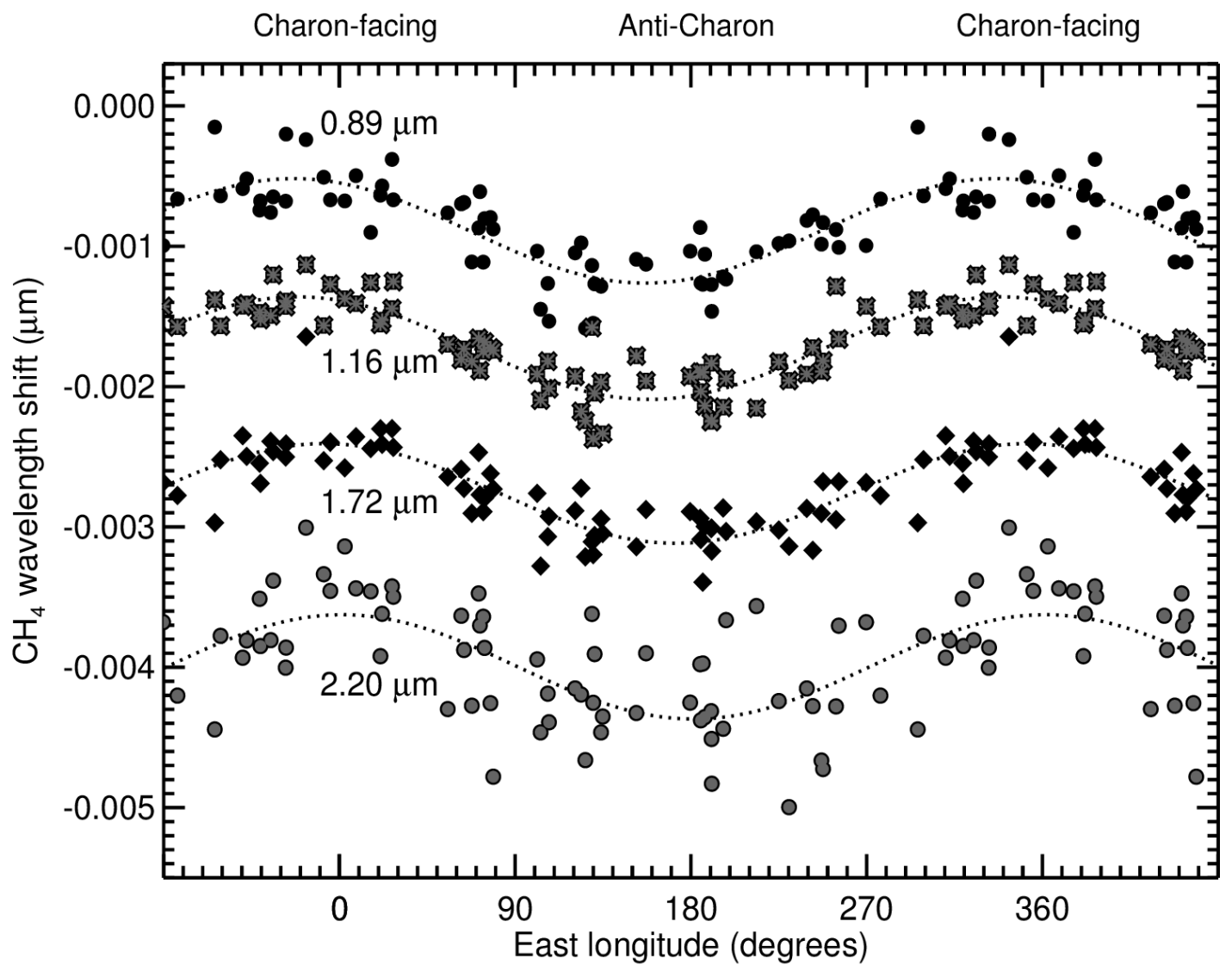

**Figure 7.** Wavelength shifts of four of Pluto's $CH_4$ bands as a function of sub-Earth longitude, as measured by cross-correlation. All bands are blue-shifted relative to the corresponding absorption bands in pure $CH_4$ (negative wavelength shifts), consistent with dilution in $N_2$ ice. In all bands, greater blue shifts are seen on the anti-Charon hemisphere, consistent with greater dilution of $CH_4$ in $N_2$ ice on that hemisphere.

# 4. Secular Evolution

The past dozen years span only a tiny fraction of Pluto's 2.5 centuries-long heliocentric orbit, but over that time the sub-Earth latitude on Pluto has progressed northward by some 20°, from around 28° N in 2001 to 48° N in 2012. It will be 51° N when the New Horizons probe encounters the system in July 2015, and will reach its maximum of about 60° N during 2025 to 2033. The recent latitude change is as large as it is because Pluto's true anomaly advances relatively rapidly near perihelion and because Pluto's obliquity is high. Secular spectral changes potentially detectable in our IRTF data could result from a static distribution of ices on Pluto with a distinct composition in more northerly latitudes increasingly contributing to the spectra due to the shifting viewing geometry. Changes could also be produced by seasonal volatile transport driving evolution of the texture, composition, or spatial distribution of Pluto's volatile ices (e.g., Hansen and Paige 1996; Spencer et al. 1997; Trafton et al. 1998; Young 2012a). The temporal baseline can be extended by including observations from the 1990s obtained at Lowell Observatory's 72" Perkins telescope (Grundy and Buie 2001), although those data have much lower spectral resolution of around $R \approx 300$ and a more restricted wavelength coverage of 1.20 to 2.35 µm. Taking the 1.72 µm $CH_4$ band as an example, its fractional depth as a function of longitude is shown in Fig. 8 for three epochs, 1998, 2001-2002, and 2011-2012. Other strong $CH_4$ bands (not shown here) exhibit similar-looking trends. Grundy and Buie (2001) observed that Pluto's strong $CH_4$ bands were getting deeper during the 1990s. That trend has evidently continued in the past decade. If the explanation is geometric, the secular increase in $CH_4$ band depths would imply strong $CH_4$ absorption at the high north latitudes on Pluto that are increasingly coming into view. The pattern of longitudinal variation has also evolved, with diurnal variability declining in amplitude consistent with the increasing contribution to the observed spectrum of northern polar regions (currently visible throughout Pluto's diurnal cycle) at the expense of more highly modulated contributions from equatorial and southern latitudes.

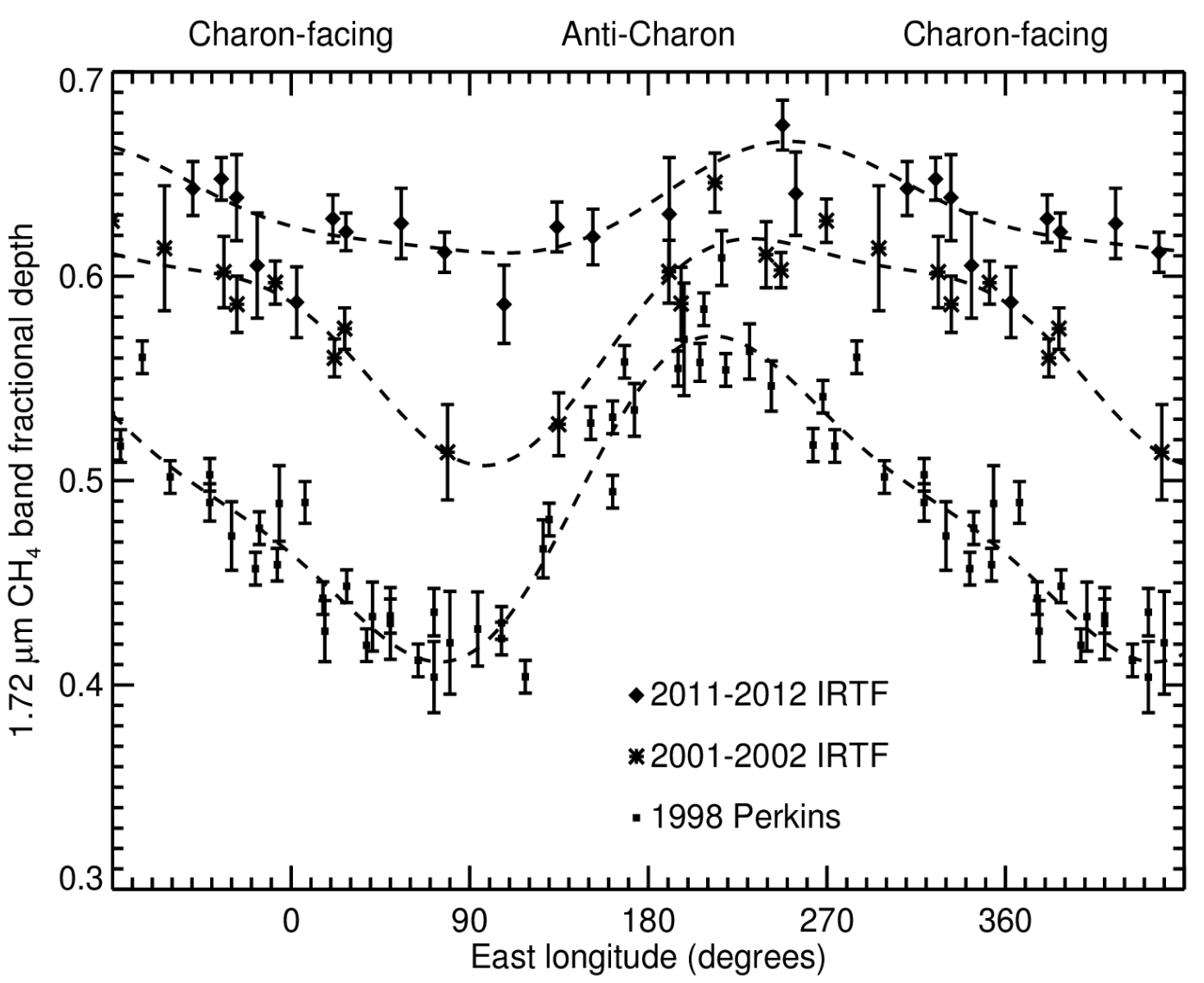

**Figure 8.** Secular evolution of the diurnal variation of $CH_4$ ice absorption on Pluto, as measured by the fractional depth of the 1.72 µm band during three different epochs, indicated by distinct plotting symbols. In creating this figure, we resampled the IRTF spectra to match the lower spectral resolution of the 1998 data. Dashed lines indicate low-order Fourier fits to the three different data subsets.

The evolution of the longitudinal variability of Pluto's strong $CH_4$ bands shares certain features with the evolution of the visible lightcurve, shown in Fig. 9 (data from Buie et al. 1997, 2010b). The amplitudes of both types of variation have decreased, with longitudes around 100° E contributing disproportionately to the decrease. Buie et al. (2010a) argued for an especially rapid period of evolution during the early 2000s, with Pluto's surface coloration and albedo patterns causing rapid changes in its lightcurve. To search for a potentially non-uniform rate of change in the $CH_4$ ice absorptions, we plotted the average band depth along with the maximum and minimum from a Fourier fit to pooled data from six different time intervals in Fig. 10. These data are also suggestive of relatively rapid evolution around the late 1990s and early 2000s.

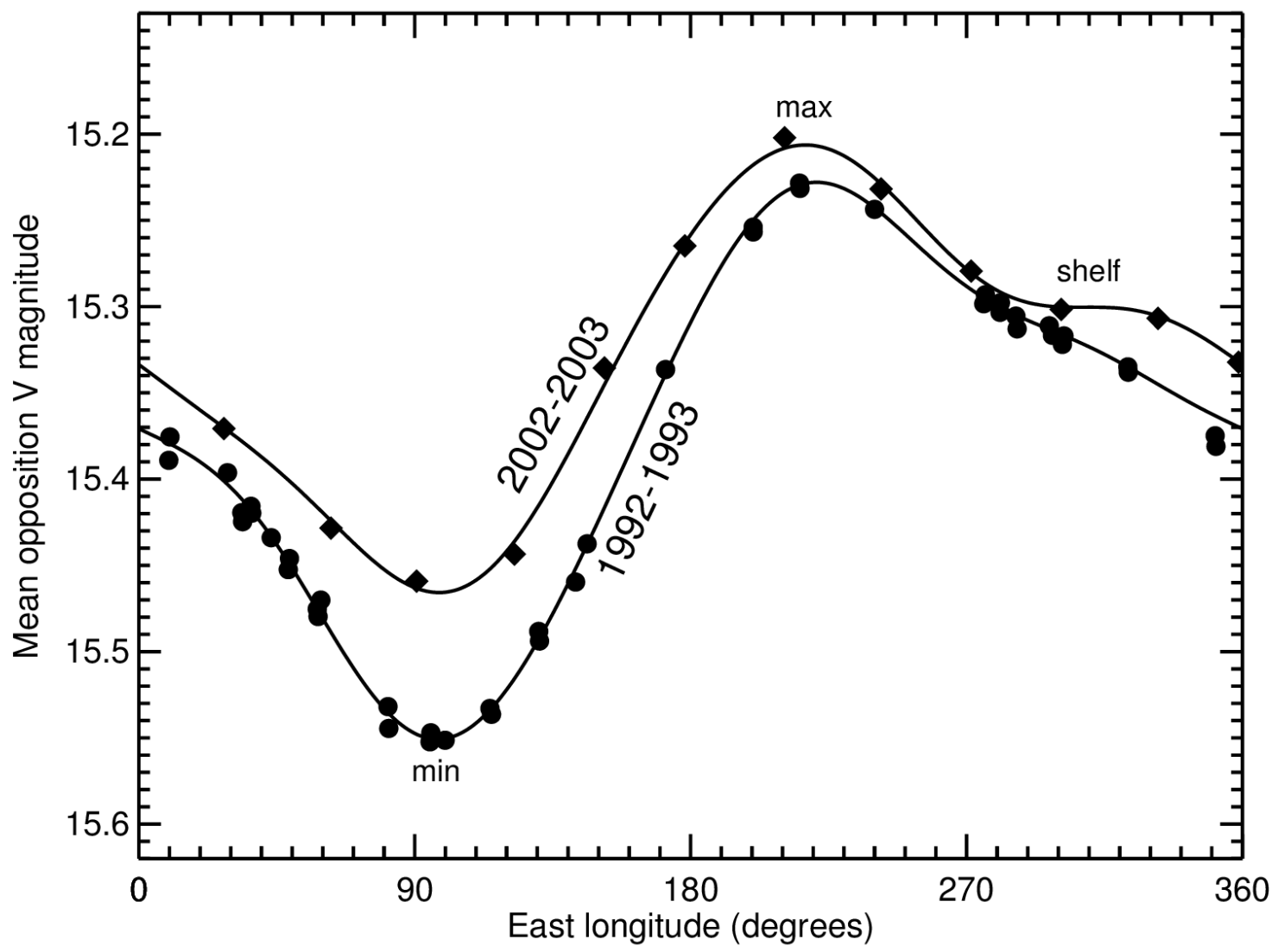

**Figure 9.** Evolution of Pluto's visible lightcurve over one decade, based on Hubble Space Telescope observations obtained during 1992-1993 and during 2002-2003 (Buie et al. 1997, 2010a; see also Buratti et al. 2003). Over that time interval, Pluto's visual lightcurve amplitude declined, mostly as a result of increasing flux near lightcurve minimum.

We noted earlier that Pluto's weak $CH_4$ absorption bands behave differently from the stronger bands, in terms of their longitudinal distribution. The same is true of their secular evolution. Grundy and Fink (1996) reported the very weak 0.72 µm band to be decreasing in strength from the 1980s through early 1990s, rather than increasing like the strong $CH_4$ bands. The weakest bands that can be reliably measured in our IRTF spectra are a little stronger than that particular band, so they may not be directly comparable, but the three weakest bands we measured do not show the secular depth increase that the stronger bands show, except right around 90° east longitude. Averaged over all longitudes, the change in $CH_4$ band depth is shown in Fig. 11 as a function of band depth for the same twelve $CH_4$ bands discussed earlier. A linear fit to these measurements becomes negative for band depths below about 10%, consistent with the Grundy and Fink (1996) observation of a declining 0.72 µm band, which had a fractional depth of around 6% at the time of those observations.

The plots of CO and $N_2$ absorption versus longitude (Figs. 3 and 6) have too much scatter to show obvious trends in just the 2001-2002 and 2011-2012 data subsets, so we grouped the data into two larger subsets, one consisting of all data taken during 2001 through 2006 (25 nights) and the other consisting of all data taken during 2007 through 2012 (39 nights). The am-plitudes and constant offsets of the sinusoidal fits shown in Figs. 3 and 6 were refitted to each six-year group of data, while holding the phase constant. For both $N_2$ and CO absorptions, the average equivalent widths declined from the first group to the second, suggesting a secular decline in absorption by both ices. The $N_2$ average dropped from 1.08 ± 0.03 to 0.82 ± 0.02 µm and the CO average dropped from 0.136 ± 0.005 to 0.125 ± 0.005 nm. The $N_2$ peak-to-peak amplitude dropped negligibly, from 0.64 ± 0.08 to 0.63 ± 0.07 µm while the CO amplitude declined from 0.153 ± 0.016 to 0.118 ± 0.015 nm. Only the change in $N_2$ average equivalent width looks statistically significant, but taken together, these numbers are suggestive of a secular decline in Pluto's CO and $N_2$ absorptions that would be consistent with those volatile ices being more prevalent at lower latitudes that are gradually becoming more foreshortened as the sub-Earth latitude moves north. If these ices were especially prevalent in the high albedo region near Pluto's anti-Charon point, we would expect to see their absorptions gradually decline as the sub-Earth latitude moves north, especially around longitude 180° E.

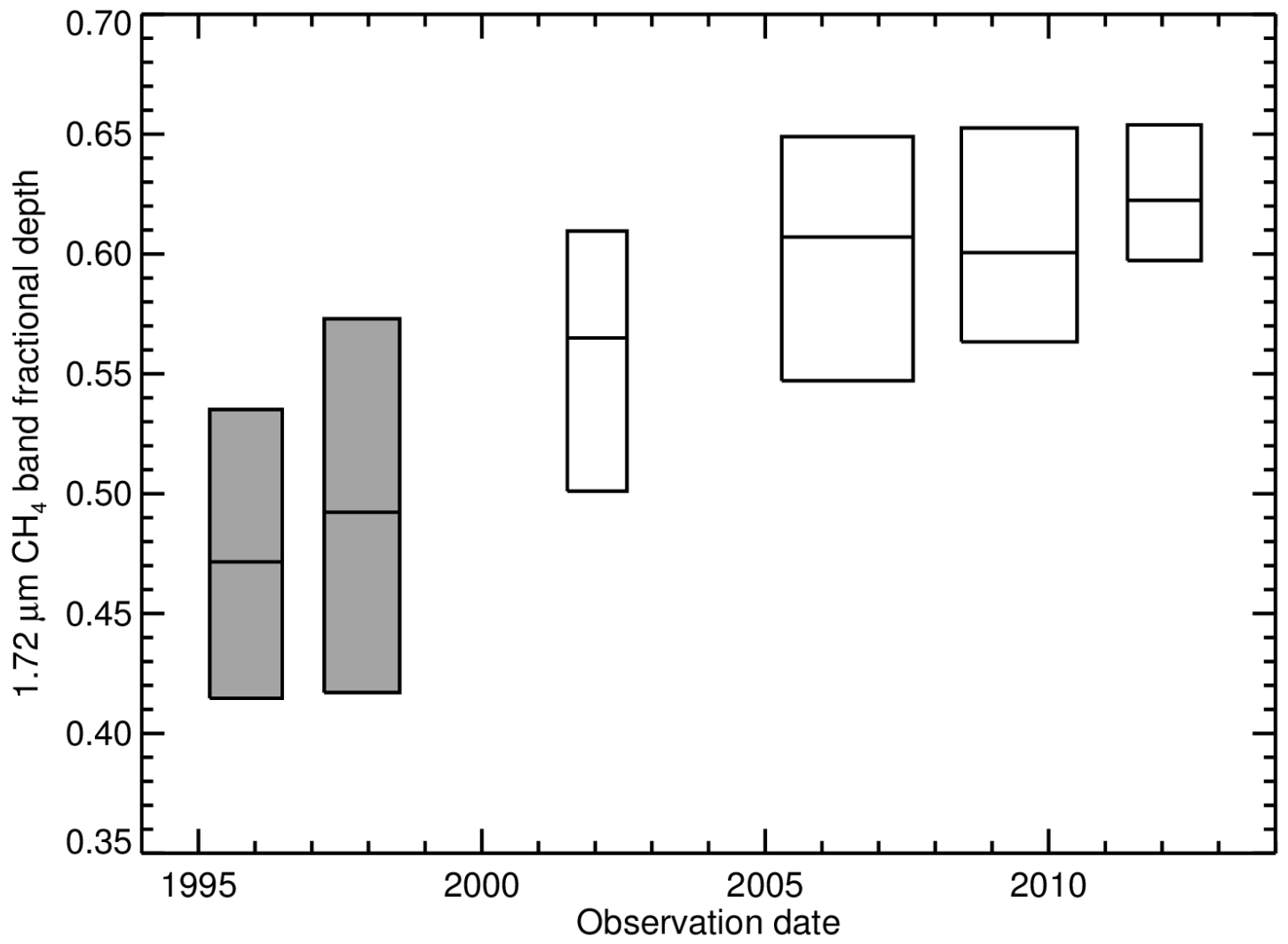

**Figure 10.** Secular evolution of Fourier fits to the longitudinal variation of $CH_4$ ice absorption, from six groups of Pluto spectra, indicated by six rectangular zones. The horizontal extent of each zone corresponds to the dates of observations included in each group. The bottom and top of each zone indicate the minimum and maximum absorption in the Fourier fit, while the central line shows the average over all longitudes. White rectangles are based on the new IRTF data presented here, while gray rectangles are based on Perkins data (Grundy and Buie 2001). The IRTF data were resampled to the lower resolution of the Perkins data in making this plot. There appears to have been somewhat more rapid evolution of the $CH_4$ ice absorption some time between ~1998 and ~2006, but the cadence of our observations limits determination of the exact time-history of this evolution (and despite our efforts to account for the different spectral resolutions, caution is always warranted when comparing data from different instruments).

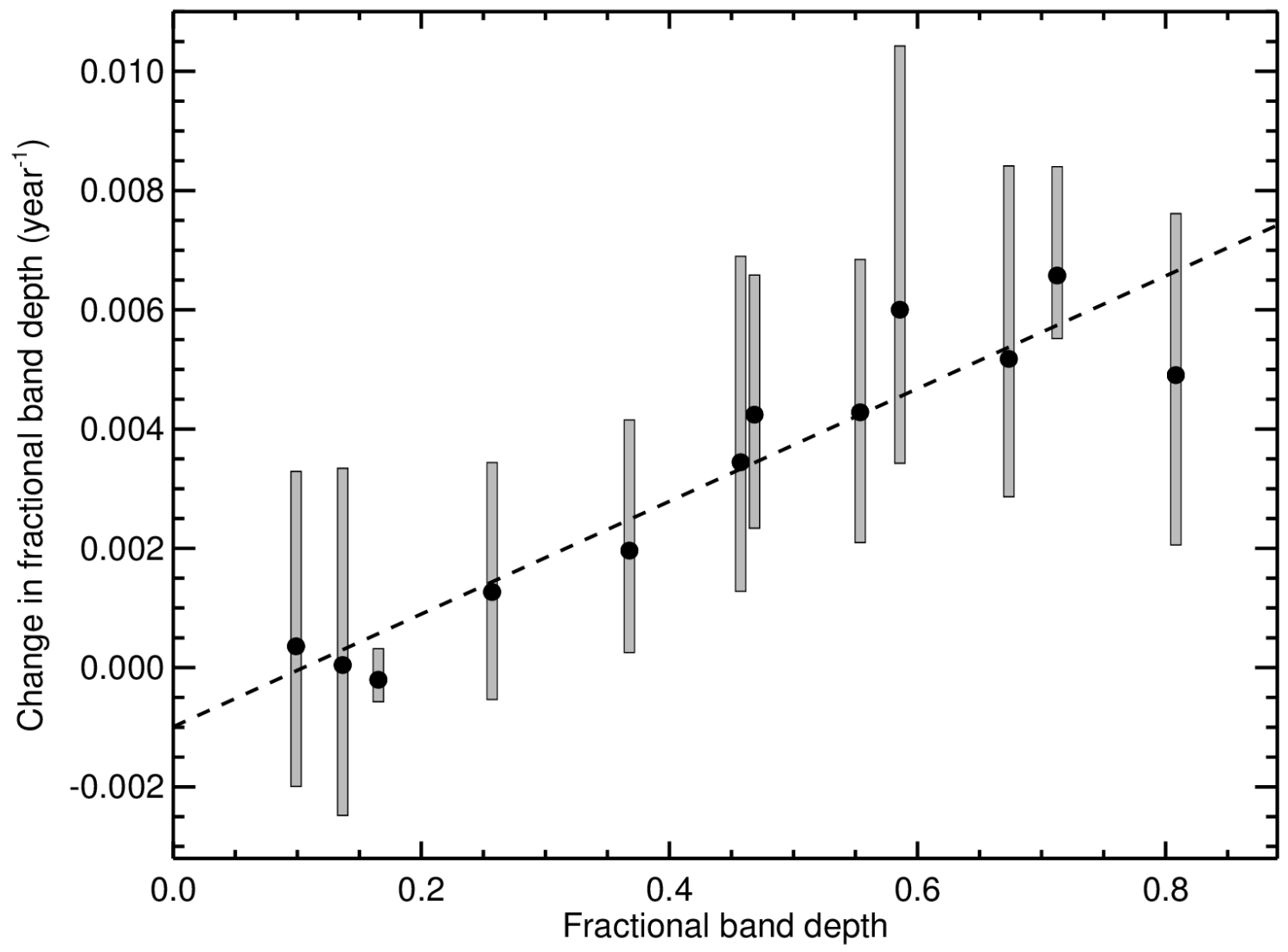

**Figure 11.** Secular changes in a dozen $CH_4$ absorption bands over the 10 year interval between 2001-2002 and 2011-2012 data subsets, showing a strong correlation between rate of change and fractional band depths. The points are averages over all longitudes from Fourier fits such as shown in Fig. 8 and the dashed line is a linear fit to those averages. The vertical bars indicate the minimum and maximum observed change for each band. The maximum change is generally seen at longitudes near 90° E while the minimum change tends to be seen around 200° E.

Another, even more exciting possibility is that we could be witnessing the seasonal removal of volatile $N_2$ and CO ices from the northern hemisphere through the action of solar-driven volatile transport. Changes in Pluto's surface ices are difficult to distinguish from changes due to geometry with our spatially unresolved observations, but Grundy and Buie (2001) suggested one possible way to identify which mechanism is responsible. Parallax from Earth's motion around the Sun offers views of Pluto differing by almost four degrees between pre- and post-opposition quadrature. Due to Pluto's high obliquity, much of the parallax translates into shifting sub-Earth latitudes on Pluto. As mentioned before, the sub-Earth latitude on Pluto has moved north by around 20° over the past dozen years, but this motion is modulated by the annual parallax variation. It is thus possible to schedule a pair of observations separated by a year plus a few months, such that the sub-Earth longitude and latitude are almost perfectly duplicated. Thanks to the impressive scheduling flexibility of the IRTF team, we were able to observe Pluto on three such pairs of UT dates, indicated as circled pairs in Fig. 2: 2005 April 15 and 2006 July 19, 2005 May 9 and 2006 July 24, and 2005 May 23 and 2006 July 25, corresponding to longitudes 105°, 188°, and 128° E, respectively. For each of these pairs, the sub-Earth longitude difference was less than 5° and the latitude difference was less than 0.5° (see Table 1). If changing geometry were solely responsible for the secular evolution, we should see little or no change in each of these matched pairs, since the observing geometries are so similar. If seasonal volatile transport were solely responsible, we would expect to see changes consistent with the longer term rate of change.

The sinusoids fitted to CO and $N_2$ ice absorptions in the 2001-2006 and 2007-2012 data subsets described above provide an estimate of the baseline rate of change in the absorptions of these ices as a function of longitude. According to the fitted sine functions, both absorptions decline over time at all three longitudes of our matched pairs. Taking the time base between the two data subsets to be six years, for the relevant longitudes we find average rates of change in equivalent widths of −0.043 nm $yr^{-1}$ for $N_2$ and −0.0038 nm $yr^{-1}$ for CO. Turning to the matched pairs, with an average time separation of 1.2 years, we see $N_2$ absorption decline by a much slower −0.004 ± 0.014 nm $yr^{-1}$ while CO absorption actually increases by +0.002 ± 0.017 nm $yr^{-1}$. Both numbers are consistent with no change at all. The $N_2$ rate of decline in the matched pair being just over 3-σ smaller than the longer term rate of −0.043 nm $yr^{-1}$ favors geometry over volatile transport as the leading explanation of the declining $N_2$ absorption over the past dozen years. But the substantial uncertainty leaves open the possibility of volatile transport accounting for at least some of the overall decline, let alone the fact that we do not know whether the decline of $N_2$ ice absorption is uniform in time, or is perhaps more punctuated or episodic, like the changes in $CH_4$ absorption and in visual colors and lightcurve seem to be. Our CO measurement in the matched pairs is unfortunately too insensitive to put comparable constraints on the cause of the apparent decline in that absorption.

Although it is less volatile than CO and $N_2$, and thus less expected to show rapid seasonal effects of volatile transport, we also examined the $CH_4$ ice bands in the geometrically matched pairs. The longer term evolution shown in Fig. 8 indicates a strong dependence on longitude that must be taken into account. That plot shows maximum secular evolution of about +0.10 µm in fractional band depth of the 1.72 µm band over ten years at a longitude of around 100° E, suggesting that of our three matched pairs, the pair obtained near 105° E should be the most sensitive to changes in $CH_4$ absorption. If volatile transport was responsible, this pair should show an increase of around +0.013 µm over the 1.3 year interval between that pair of observations. Instead we observed a decrease of −0.018 ± 0.010 µm. Using the 2001-2002 and 2011-2012 data subsets, we measured the changes for the 12 $CH_4$ bands in Table 2. Figure 11 shows larger long-term increases for the stronger bands. We did the same for our 105° E matched pair, and only the three weakest bands showed increases. The strong bands weakened. We would have expected a strengthening trend in the matched pairs if volatile transport alone explained the observed long-term trend, and no change if geometry alone was responsible. Neither hypothesis called for a weakening of these bands. Grundy and Buie (2001) noted a similar weakening trend in geometrically matched pairs from 1997 and 1998, although it was not statistically distinguishable from zero in their data. If we interpret the 1.72 µm band data as saying a −0.018 µm change in band depths in 1.3 years was due to volatile transport, that would imply an offsetting change of +0.031 µm in 1.3 years due to geometric factors, in order to reconcile the two observations. Such a scenario is possible if a $CH_4$-rich northern polar cap is rotating into view and losing its $CH_4$ ice absorptions to volatile transport at the same time. But again, potential non-uniformity of the secular change suggested by Fig. 10 raises questions as to the validity of the test offered by the geometrically matched pairs.

Other possible types of change could also affect the observed spectra. For instance, changing ice textures could lead to observable changes in band depth. A shrinking mean grain size in a particulate surface, perhaps due to sublimation or fracturing, would lead to reduced optical path lengths and thus reduced absorption band depths. Scenarios like this are to be explored in a separate paper involving radiative transfer modeling of the absorption bands. A variety of shorter term transient changes could also potentially be detected in time-series spectral data. These might occur in the wake of some sort of large scale eruption or weather event, as has been suggested for Triton (Hicks and Buratti 2004). Outlier points, the sort of signature that might be expected from a transient event, can certainly be found in our plots of various absorption band strengths versus longitude. But in all cases, the outliers were found to arise from the lowest quality spectra obtained, attributable to poor sky conditions and/or short total integration times. Accordingly, it seems unwise to ascribe any significance to them.

# 5. Grand Average

Combining data from all 65 nights yields the very high signal-to-noise grand average spectrum shown in Fig. 12. Since the wavelengths of SpeX's pixels shift slightly from night to night, merging the spectra required resampling them all to a common wavelength scale, for which we used a wavelength grid with a constant resolution of $R$ = 1100 (wavelength divided by double the sample spacing), providing a reasonable match to our lower resolution spectra. Detailed radiative transfer modeling of this grand average product will be presented in companion papers, but some quick comparisons are worth making here.

An examination of the shape of the $N_2$ ice absorption at 2.15 µm shows it to be consistent with the hexagonal $\beta$ phase of $N_2$ ice (see Fig. 13). The cubic $\alpha$ phase of $N_2$ has a much narrower band. That phase is only stable at temperatures below 35.61 K for pure $N_2$ ice (Scott 1976), es-tablishing a firm lower limit on Pluto's $N_2$ ice temperature. A much cruder upper limit on the $N_2$ ice temperature can be set by the presence of a shoulder on the right side of the $N_2$ band at around 2.162 µm. In pure $N_2$ ice, this feature only appears below approximately 41 K (Grundy et al. 1993), implying an $N_2$ ice temperature between 35.61 and ~41 K, consistent with the conclusions of Tryka et al. (1994). However, caution is advisable for two reasons. First, published correspondences between spectral features and temperatures are for pure $N_2$ ice, and as discussed earlier, Pluto's $N_2$ ice should have $CH_4$ and CO dissolved in it (and perhaps also argon, e.g., Tegler et al. 2010). More laboratory work is needed on the temperature-dependent spectral behavior of $N_2$ in the presence of these probable contaminants. Second, the hydrogen Brackett *γ* line (Brackett 1922) produces a Fraunhofer absorption in the solar spectrum at 2.160 µm. We checked ratios between spectra of our different solar analog stars and this line seems to cancel out reasonably well in the ratios, but we do not have a direct comparison between the solar analogs and the Sun itself. Flexure within the instrument can result in wavelength shifts and thus imperfect cancellation, even if the stellar emission perfectly matched the solar spectrum.

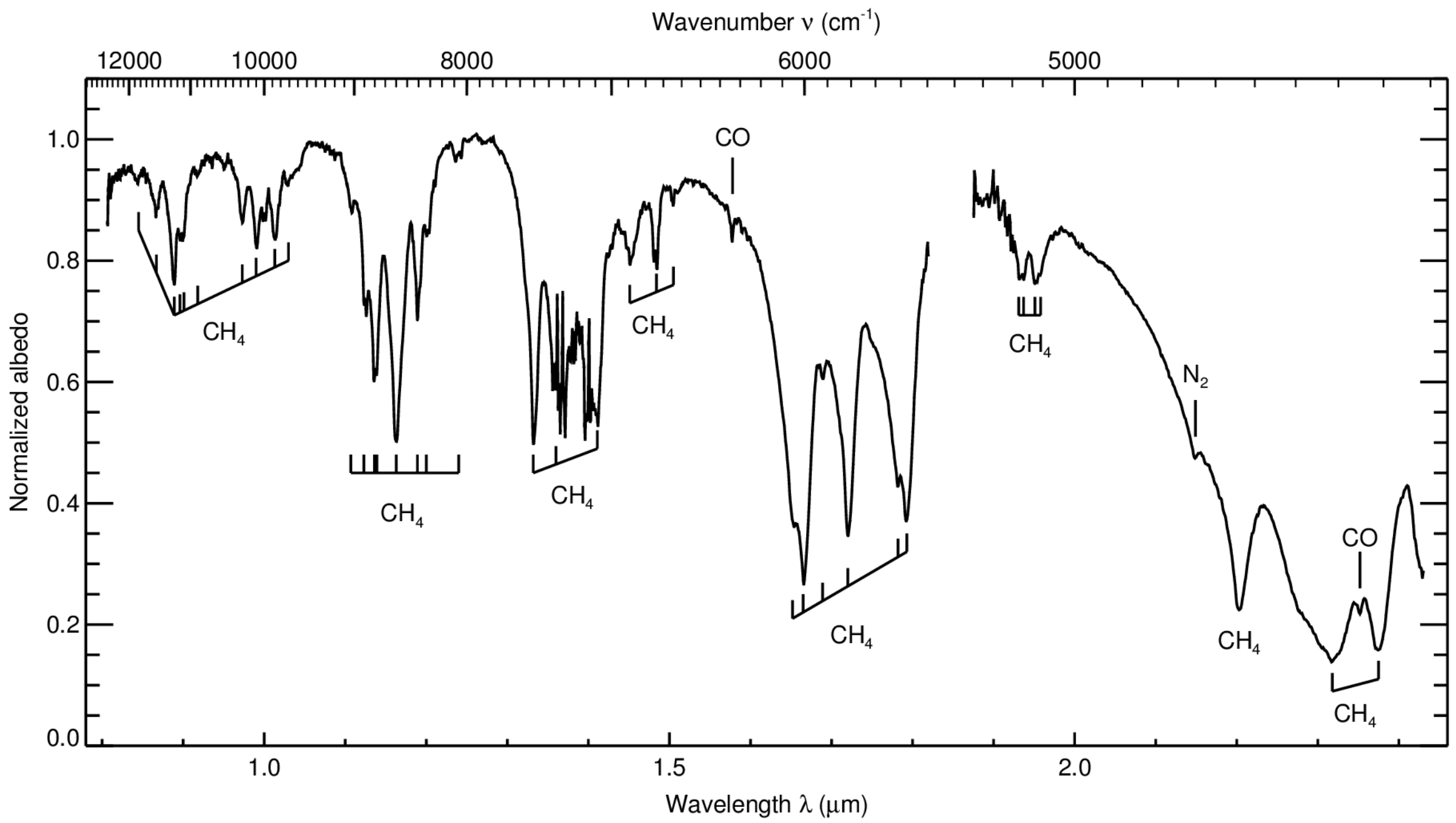

**Figure 12.** Grand average IRTF/SpeX spectrum of Pluto with absorptions of CO, $CH_4$, and $N_2$ ices labeled.

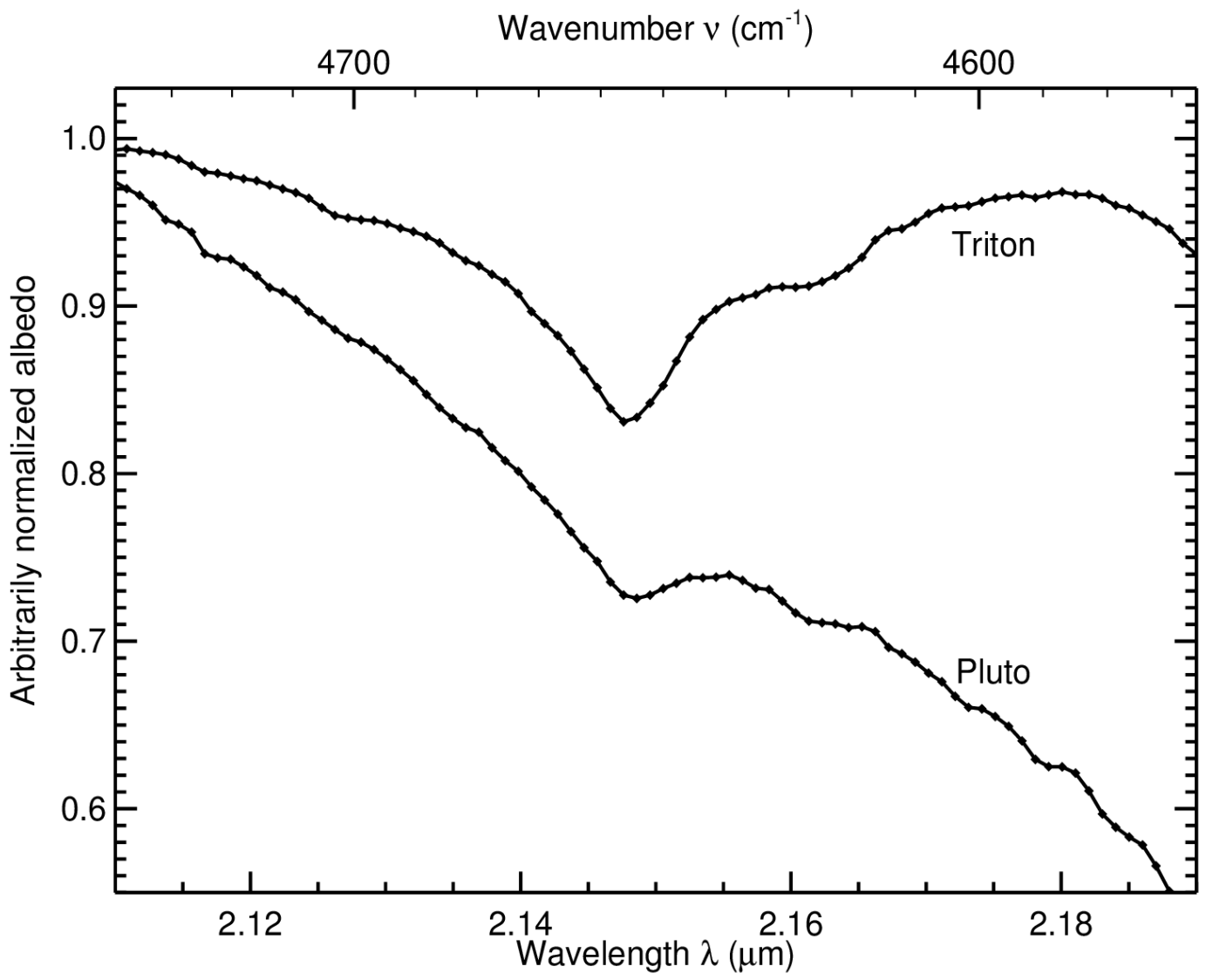

**Figure 13.** Comparison of our Pluto grand average spectrum with that of Triton (from Grundy et al. 2010) in the region of the $N_2$ ice overtone band, showing the main $N_2$ absorption at 2.147 µm and also a weaker side-band at about 2.162 µm, indicative of $N_2$ ice temperatures below ~41 K.

In spectra of Triton, Neptune's large captured satellite, $CO_2$ ice has been detected through three strong, narrow, vibrational absorption bands near 2 µm, along with several weaker bands (Cruikshank et al. 1993, 1998b; Hansen 1997; Grundy et al. 2010). These absorptions have not been seen in spectra of Pluto. Figure 14 shows a zoomed-in view of both Triton and Pluto grand averages in the region of the $CO_2$ ice bands. No trace of $CO_2$ ice absorption can be seen in the Pluto spectrum. Not being detected at the signal precision of our grand average spectrum implies that less than 3% of the observed surface of the Pluto-Charon system could be covered with Triton-like material. DeMeo et al. (2010) searched for ethane ice absorptions in Pluto spectra obtained in 2005, finding features consistent with pure $C_2H_6$ ice at 2.274, 2.405, 2.457, and 2.461 µm. Our grand average spectrum looks consistent with their data at 2.274 and 2.405 µm (see Fig. 12), but our wavelength coverage does not extend to the two longer wavelength bands. DeMeo et al. also searched for, but did not see a weaker $C_2H_6$ ice absorption at 2.015 µm in their Pluto spectra. Our grand average also shows no obvious absorption at that wavelength, nor at the wavelengths of three other nearby $C_2H_6$ ice bands at 2.021, 2.030, and 2.054 µm.

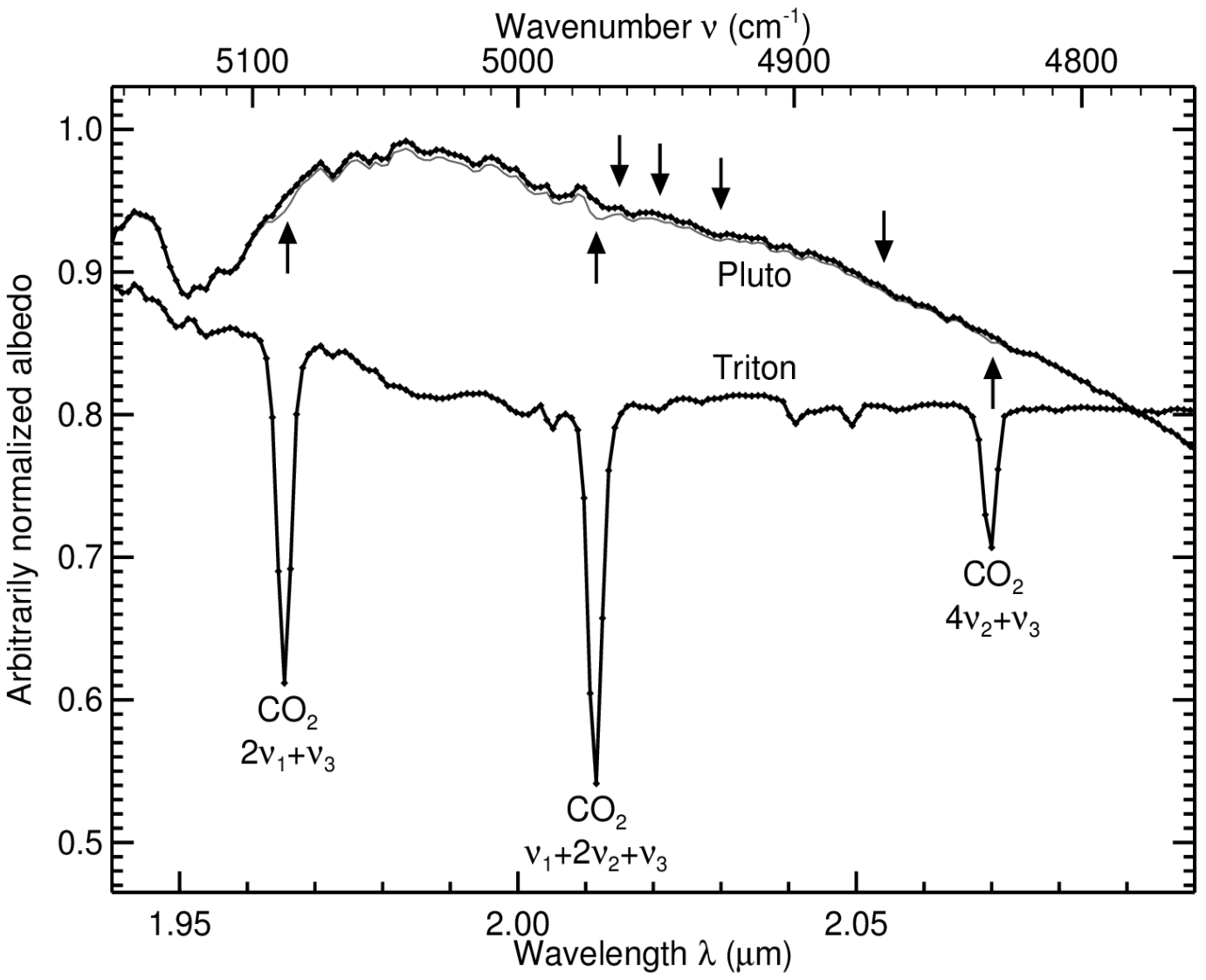

**Figure 14.** Comparison of our Pluto grand average spectrum with that of Triton (from Grundy et al. 2010) in the region where Triton's spectrum features three strong and narrow $CO_2$ ice absorption bands. No trace whatsoever of these $CO_2$ ice absorptions can be seen in the Pluto spectrum at the relevant wavelengths, indicated by upward-pointing arrows. The thin gray curve is a linear mixture of 3% Triton and 97% Pluto. If 3% of Pluto's surface had been composed of Triton-like material, the $CO_2$ absorptions would have been visible in our data. The downward-pointing arrows indicate wavelengths of $C_2H_6$ ice absorption bands. Corresponding dips are not apparent in our Pluto grand average spectrum.

## 6. Summary

New 0.8-2.4 µm spectral observations of the Pluto system were obtained at IRTF/SpeX during 65 nights from 2001 through 2012. The data reveal striking inhomogeneities in the spatial distributions of Pluto's ice absorptions. When the anti-Charon hemisphere (180° E) is oriented toward Earth, carbon monoxide ice absorption is almost triple what is seen half a Pluto day later, when the sub-Charon hemisphere (0° E) is oriented toward Earth. Although there is considerable scatter in our $N_2$ ice equivalent width numbers, the longitudinal pattern of that absorption appears to be at least crudely similar to that of the CO ice. Blue shifts of $CH_4$ bands, attributed to dilution of $CH_4$ in $N_2$ ice, are also at their greatest at this longitude. In contrast, the strong $CH_4$ ice bands show their maximum absorption around 260° E, almost 90° out of phase with the CO and $N_2$ ices. Weaker $CH_4$ absorptions are even further out of phase with CO and $N_2$, with their maximum absorptions shifted even further toward the sub-Charon hemisphere. These longitudinal patterns are illustrated schematically in Fig. 15. Grundy and Fink (1996) divided an albedo map of Pluto into three zones according to albedo to create a hypothetical distribution of three different compositional provinces on Pluto. This distribution of ices was able to match a number of features of Pluto's spectral variability and has subsequently been adapted and modified by various authors (e.g., Lellouch et al. 2000, 2011; Grundy and Buie 2001). The new IRTF data set offers an important test for the various versions of this map, to be presented in a companion paper.

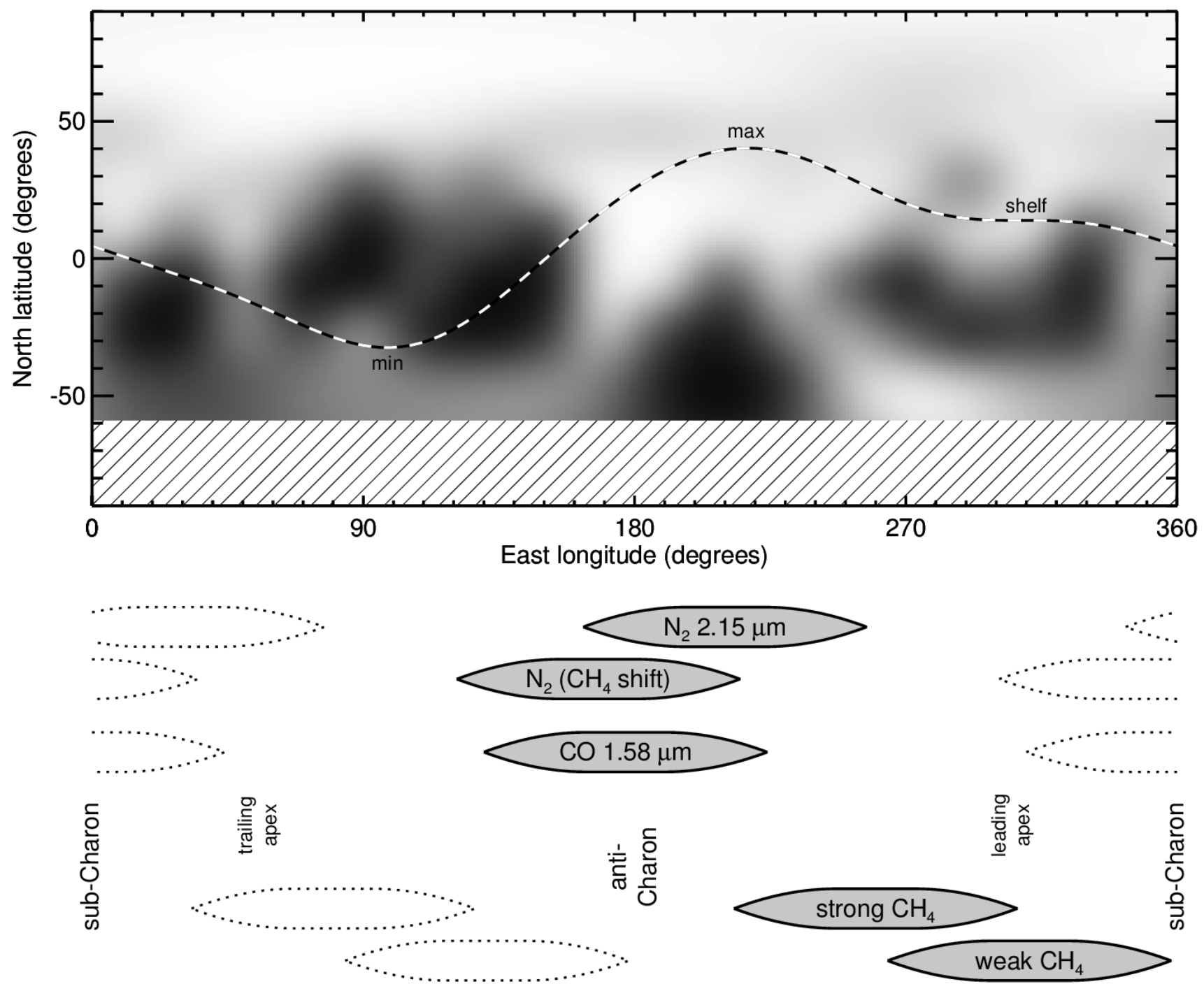

**Figure 15.** Approximate longitudes where various spectral characteristics reach their maxima in sinusoidal fits are indicated by gray battleship hulls, with corresponding minima indicated by dotted outlines at the same height. Also shown are the Buie et al. (2010a,b) lightcurve and *V* wavelength albedo map from HST observations during 2002-2003. In addition to sub- and anti-Charon longitudes on Pluto, the leading and trailing apex longitudes are marked. These are offset from the usual 270° and 90° for a tidally-locked satellite due to Pluto's proximity to the system barycenter.

$CH_4$ ice is much less volatile than $N_2$ and CO, with a vapor pressure at 40 K roughly three orders of magnitude below those of CO and $N_2$ (Fray and Schmitt 2009). The volatility contrast between $CH_4$ and the other two ices suggests that their distinct distributions could represent the product of sorting by volatility. Such a pattern could perhaps be controlled by elevation if Pluto has a fossil tidal bulge or a displacement of its center of mass relative to its center of body, leading to a near-side/far-side dichotomy like Earth's moon has. Large-scale patterns in Pluto's internal heat flow offer another possible explanation. The remarkably unbalanced volatile ice distribution is reminiscent of the Moore and Spencer (1990) “Koyaanismuuyaw” hypothesis of a perennially dichotomous Triton. The Charon-centered symmetry of Pluto's $N_2$ and CO ice distributions is especially intriguing. Triton's volatile ices show a comparably symmetric distribution around Triton's sub-Neptune point, except that in Triton's case the $N_2$ and CO ices are concentrated on the sub- rather than the anti-Neptune hemisphere (Grundy et al. 2010).

Longer term changes in Pluto's ice absorptions are also evident in the data. The strong $CH_4$ bands are getting stronger while the CO and $N_2$ absorptions are declining in strength. A possible explanation could be that the north polar regions that have been swinging more into view since equinox in the mid-1980s have more $CH_4$ ice and less CO and $N_2$. However, volatile transport models (e.g., Hansen and Paige 1996; Spencer et al. 1997; Trafton et al. 1998; Young 2012a) suggest that volatile $N_2$ (and presumably also CO) should have been accumulating at high northern latitudes during the long southern summer that just ended in the 1980s. Now that those northern latitudes are receiving continuous sunlight, the CO and $N_2$ should be sublimating away, but the fact that the atmospheric pressure is increasing rather than declining suggests they have not yet been exhausted from the northern hemisphere (Young 2012a,b). To get an idea of whether the observed decline in $N_2$ and CO is due to sublimation removal of those species from observable latitudes on Pluto, or simply from the changing observing geometry, we did observations on three pairs of nights selected to be separated in time by a little over a year, but with almost no change in observing geometry. Those observations showed less change than would be expected on the basis of the time interval between them, and thus favor the geometric explanation. Oddly, the $CH_4$ absorptions in the geometrically matched pairs diminished, contrary to the long-term strengthening trend for those absorptions. Both long-term and matched-pair $CH_4$ behaviors can be explained if volatile transport is reducing the strengths of the $CH_4$ bands at the same time as geometric effects are increasing them. Averaged over the past decade, the two effects combine to result in an overall increase in $CH_4$ band strengths, but if geometry is held constant, only the volatile transport weakening effect is seen.

This evidence for complex and time-variable spatial distributions of volatile ices on Pluto is tantalizing, but ambiguities between geometric and volatile transport interpretations of spectral changes hamper interpretation of the behavior of Pluto's volatile ices. NASA's New Horizons spacecraft will get a much more detailed view when it flies through the Pluto system in 2015 (Young et al. 2008). New Horizons' Linear Etalon Imaging Spectral Array (LEISA) will return near-infrared spectral maps covering wavelengths from 1.25 to 2.50 µm at a spectral resolution of $R \approx 240$ with higher resolution coverage of $R \approx 560$ between 2.10 and 2.25 m (Reuter et al. 2008). LEISA will map Pluto's ices at spatial resolutions as high as 10 km/pixel. Once the 2015 distribution of ices is known, ambiguities between geometric and volatile transport effects can be broken, making a long-term spectral monitoring data set such as presented here immensely valuable for investigating the rate at which Pluto's spatial distributions of ices are changing in time. However, we should also emphasize the need for laboratory measurements currently in progress on temperature-dependent spectral behaviors of the mixed $N_2$, CO, and $CH_4$ phases expected to be present on the surface of Pluto from thermodynamic considerations. These data are essential for interpretations of the New Horizons data, as well as ground-based spectral observations such as presented here.

## Acknowledgments

The authors gratefully thank the staff of IRTF for their tremendous assistance over the dozen years of this project, especially W. Golisch, D. Griep, P. Sears, E. Volquardsen, S.J. Bus, J.T. Rayner, and A.T. Tokunaga. As with any long-term project, a great many other people and organizations have contributed in various ways over the years, including R.S. Bussmann, K. Crane, J.R. Spencer, J.A. Stansberry, S.B. Porter, and S.D. Tegler, and JPL's Horizons ephemeris services. The work was funded in part by NSF grants AST-0407214 and AST-0085614 and NASA grants NAG5-4210 and NAG5-12516. We wish to recognize and acknowledge the significant cultural role and reverence of the summit of Mauna Kea within the indigenous Hawaiian community and to express our appreciation for the opportunity to observe from this special mountain. Finally, we thank the free and open source software communities for empowering us with key software tools used to complete this project, notably Linux, the GNU tools, LibreOffice, MySQL, Evolution, Python, and FVWM.